\documentclass{aa}  

\usepackage{graphicx}
\usepackage{txfonts}
\usepackage{xcolor}
\usepackage{ulem}
\usepackage{url}
\usepackage{longtable}
\usepackage{multirow}
\usepackage{siunitx}
\usepackage{booktabs}
\usepackage{amsmath}
\begin{document}

   \title{The nature of point source fringes in mid-infrared spectra acquired with the James Webb Space Telescope \thanks{Example code and more technical information can be found via: \protect\url{https://github.com/YannisArgyriou/miri_mrs_point_source_fringes.git}}}

   \author{Ioannis Argyriou\inst{1}
          \and
          Martyn Wells\inst{2}
          \and
          Alistair Glasse\inst{2}
          \and
          David Lee\inst{2}
          \and
          Pierre Royer\inst{1}
          \and
          Bart Vandenbussche\inst{1}
          \and 
          Eliot Malumuth\inst{3}
          \and
          Adrian Glauser\inst{4}
          \and
          Patrick J. Kavanagh\inst{5}
          \and
          Alvaro Labiano\inst{6}
          \and
          Fred Lahuis\inst{7}
          \and
          Michael Mueller\inst{8,7}
          \and
          Polychronis Patapis\inst{4}
         }

   \institute{Instituut voor Sterrenkunde, KU Leuven,
              Celestijnenlaan 200D, bus-2410, 3000 Leuven, Belgium\\
              \email{ioannis.argyriou@kuleuven.be}
              \and
              UK Astronomy Technology Centre, Royal Observatory, Blackford Hill Edinburgh, EH9 3HJ, Scotland, United Kingdom
              \and 
              Telophase Corporation/Code 667, NASA’s Goddard Space Flight Center, Greenbelt, MD 20771
              \and 
              Institute of Particle Physics and Astrophysics, ETH Zurich, Wolfgang-Pauli-Str 27, 8093 Zurich Switzerland
              \and
              School of Cosmic Physics, Dublin Institute for Advanced Studies, 31 Fitzwilliam Place, Dublin 2, Ireland
              \and
              Centro de Astrobiología (CAB, CSIC-INTA), ESAC - Camino Bajo del Castillo s/n, 28692 Villanueva de la Cañada (Madrid)
              \and
              SRON Netherlands Institute for Space Research, P.O. Box 800, 9700 AV Groningen, The Netherlands
              \and
              Sterrewacht Leiden, P.O. Box 9513, 2300 RA Leiden, The Netherlands
         }

   \date{Received July 31, 2020}

 
  \abstract
    {As is common for infrared spectrometers, the constructive and destructive interference in different layers of the James Webb Space Telescope (JWST) Mid-Infrared Instrument (MIRI) detector arrays modulate the detected signal as a function of wavelength. The resulting "fringing" in the Medium-Resolution Spectrometer (MRS) spectra varies in amplitude between 10\%\ and 30\%\ of the spectral baseline. A common method for correcting for fringes relies on dividing the data by a fringe flat. In the case of MIRI MRS, the fringe flat is derived from measurements of an extended, spatially homogeneous source acquired during the thermal-vacuum ground verification of the instrument. While this approach reduces fringe amplitudes of extended sources below the percent level, at the detector level, point source fringe residuals vary in a systematic way across the point spread function (PSF). The effect could hamper the scientific interpretation of MRS observations of unresolved sources, semi-extended sources, and point sources in crowded fields.}
   {We find MIRI MRS point source fringes to be reproducible under similar observing conditions. We want to investigate whether a generic and accurate correction can be determined. Therefore, we want to identify the variables, if they exist, that would allow for a parametrization of the signal variations induced by point source fringe modulations.}
   {We determine the point source fringe properties by analyzing MRS detector plane images acquired on the ground. We extracted the fringe profile of multiple point source observations and studied the amplitude and phase of the fringes as a function of field position and pixel sampling of the point spread function of the optical chain.}
   {A systematic variation in the amplitude and phase of the point source fringes is found over the wavelength range covered by the test sources ($4.9-5.8$ $\mu m$). The variation depends on the fraction of the point spread function seen by the detector pixel. We identify the non-uniform pixel illumination as the root cause of the reported systematic variation. This new finding allows us to reconcile the point source and extended source fringe patterns observed in test data during ground verification. We report an improvement after correction of 50$\%$ on the 1$\sigma$ standard deviation of the spectral continuum. A 50$\%$ improvement is also reported in line sensitivity for a benchmark test with a spectral continuum of 100 mJy. The improvement in the shape of weak lines is illustrated using a T Tauri model spectrum. Consequently, we verify that fringes of extended sources and potentially semi-extended sources and crowded fields can be simulated by combining multiple point source fringe transmissions. Furthermore, we discuss the applicability of this novel fringe-correction method to the MRS data (and the data of other instruments).}
   {}

               
  \titlerunning{The nature of point source fringes in mid-infrared spectra acquired with MIRI onboard JWST}

  \authorrunning{I. Argyriou et al.}

  \maketitle
%

\section{Introduction}
   The James Webb Space Telescope (JWST) is an international collaboration between NASA, the European Space Agency (ESA), and the Canadian Space Agency (CSA), planned for launch in 2021. The space telescope will be placed in a halo orbit around the Lagrangian L2 point. It has a segmented primary mirror 6.5 meters in diameter and it is equipped with a large sun-shield to protect its instruments from the strong infrared light from the Sun, Earth, and Moon. From its L2 orbit, JWST will observe the light from distant astronomical objects in the near and mid-infrared \citep{jwst1}, using four scientific instruments: the Near-Infrared Imager and Slitless Spectrograph \citep[FGS/NIRISS,][]{niriss}, Near-Infrared Camera \citep[NIRCam,][]{nircam}, Near-Infrared Spectrograph \citep[NIRSpec,][]{nirspec}, and Mid-Infrared Instrument \citep[MIRI,][]{miri_pasp_1}.
   
   The mid-infrared instrument MIRI has four operational modes: (1) imaging, (2) coronagraphy, (3) low-resolution spectroscopy, and (4) medium-resolution spectroscopy \citep{miri_pasp_2,miri_pasp_3,miri_pasp_4,miri_pasp_5}. Focusing on the MIRI medium-resolution spectrometer (MRS), this is an integral-field spectrograph \citep[IFS,][]{Wells_2015}. The spectrograph contains four integral field units (IFUs) with a FOV of 3.9$\times$3.9~arcseconds to 7.7$\times$7.7~arcseconds. Each IFU records a 2D spectrum of the field of view (FOV) with a resolution ($\lambda/\Delta\lambda$) of 3500 to 1500 in the wavelength range of $4.9-28.8$~$\mu m$. The spectra are then recombined to form a 3D spectral cube.
   
   Fringing in the detector arrays of infrared spectrographs, such as the MIRI MRS, has long been known to be an issue. Constructive and destructive interference in different layers of the detector arrays modulate the detected signal as a function of wavelength. The Space Telescope Imaging Spectrograph (STIS) onboard the \textit{Hubble Space Telescope}, operating in the near-infrared, is also impacted by the same phenomenon, despite using different detectors optimized for both the ultraviolet and infrared. The amplitude, frequency, and phase of the fringes depend on the refractive and geometric properties of the detector, as shown in \citet{stis_fringing_malumuth}. This dependency was observed for the MIRI MRS as soon as the first test-data were acquired.
   
    Understanding how fringes manifest across the MRS detectors, and crucially how these depend on the spatial profile of a source, is important for correctly characterizing the spectral modulation and minimizing the uncertainty in the science performed with the MRS. This applies to point sources and, furthermore, to extended sources with or without spatial structure. To study the fringes in MIRI MRS spectra, we use ground verification data from the JWST Integrated Science Instruments Module (ISIM) cryo-vacuum (CV) tests. The JWST ISIM module, holding all JWST science instruments, was tested under operational temperature conditions during the CV tests and was fed with optical stimuli, simulating the JWST telescope optics. 
    
    Our paper is structured as follows. In Sect.~\ref{sec:instrument_description}, we provide a short description of the operating principles of the MIRI MRS, the type of detector used, and how fringes manifest in the MRS detectors. In Sect.~\ref{sec:spectral_extraction}, we describe the data and the methodology used to derive the fringe transmission from MRS detector plane images. In Sects. \ref{sec:fringe_amplitude} and \ref{sec:fringe_phase}, we analyze all available point source data using the methodology described in Sect.~\ref{sec:spectral_extraction}. The global trends in fringe depth and phase across all observations are shown therein. In Sect.~\ref{sec:point_source_fringe_correction}, we use the new fringe information to derive a new point source fringe correction. We examine the impact of the new correction on the spectral continuum determination, line sensitivity, and shape of weak spectral lines in Sect.~\ref{sec:point_source_fringe_correction_improvement}. In Sect.~\ref{sec:from_point_to_extended}. we simulate an extended source spectrum by combining multiple MIRI MRS point source spectra. We verify that this simulated extended source spectrum has an equivalent fringe profile that matches that of a real extended source observed with MIRI. Based on these results,  a fringe correction procedure is presented for sources with spatial structure and for crowded fields in Sect.~\ref{sec:semi_extended_source_correction}. Finally, we present the conclusions of our study in Sect.~\ref{sec:conclusions}.
    
   \section{Instrument description}
   \label{sec:instrument_description}
   
   \subsection{The MIRI Medium-Resolution Spectrometer}
\begin{figure}
   \includegraphics[height=5.2cm]{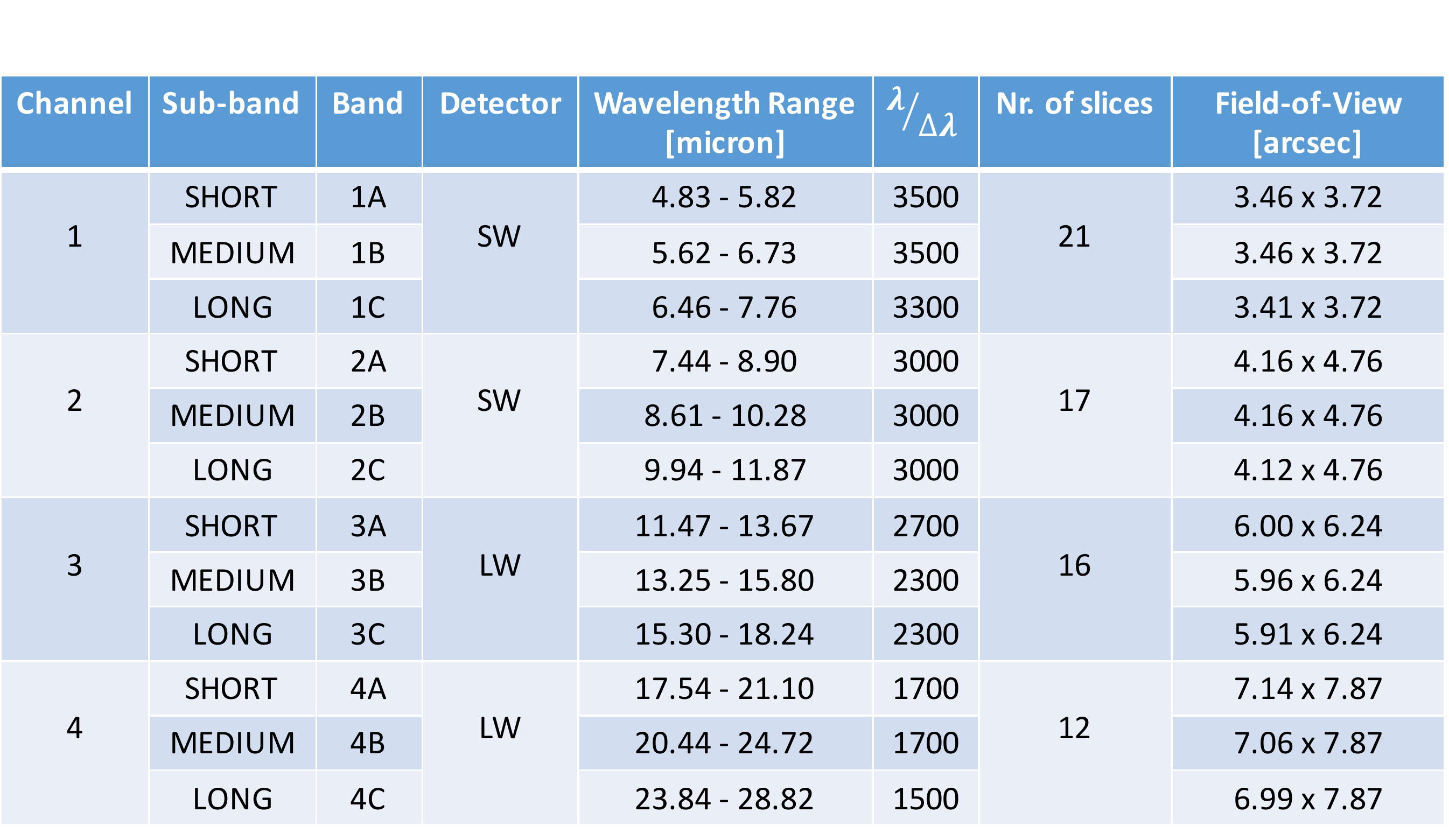}
   \caption[]
   { \label{fig:mrs_aux}
   Reference values for each MIRI MRS spectral band.}
   \end{figure}

\begin{figure*}
   \centering
   \includegraphics[height=9cm]{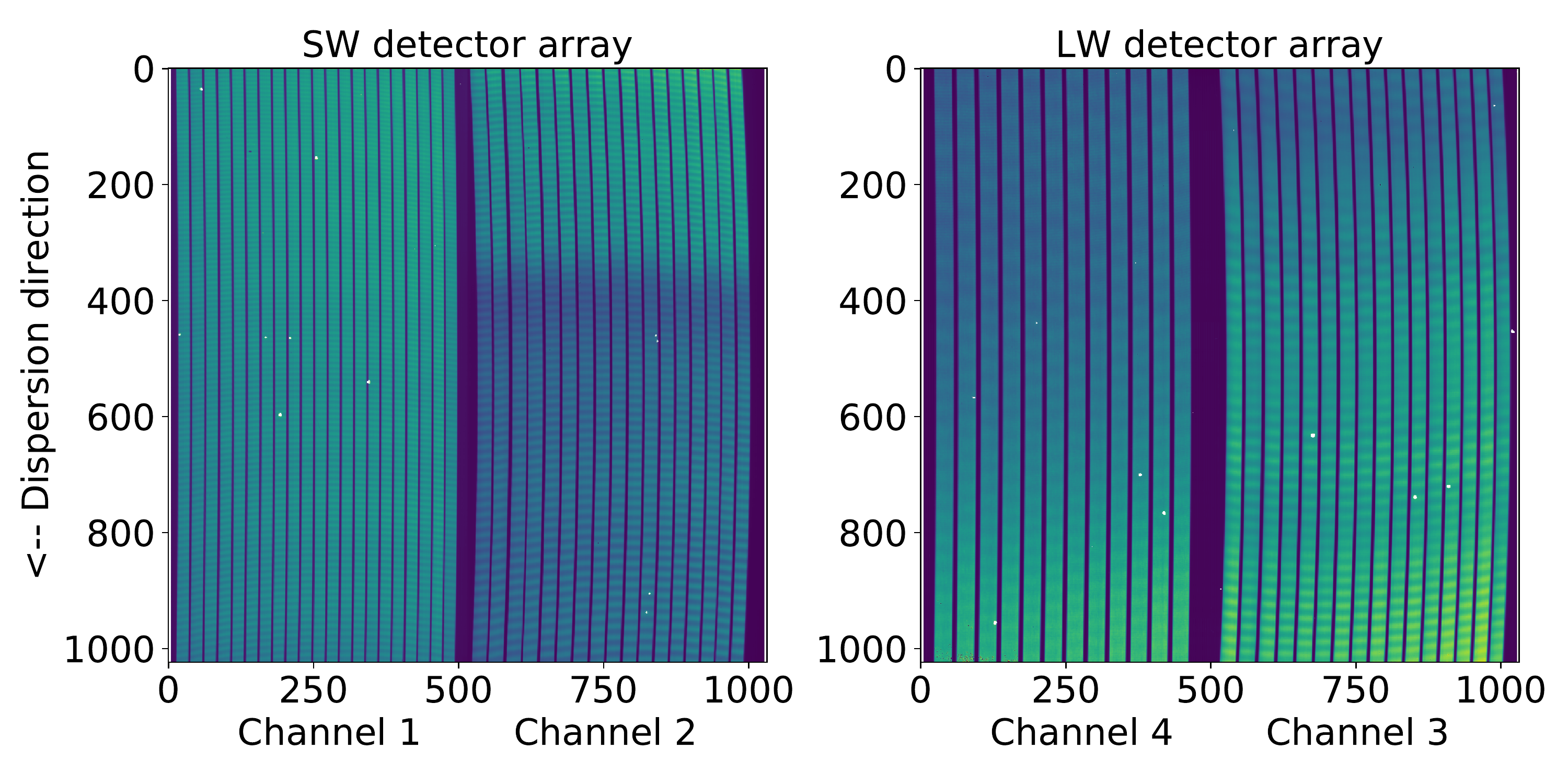}
   \caption[]
   { \label{fig:allbands_detimg}
   Spatially uniform source observed with the MIRI MRS on-ground. The MIRI MRS optics split the wavelength range into four wavelength channels. Each channel is dispersed on a different half of the MRS short-wavelength and long-wavelength detectors (labeled SW and LW detector arrays respectively). The alternating bright and dark stripes in every slice are due to fringing.}
   \end{figure*}

   The light entering the MIRI MRS is spectrally separated into four channels by dichroic filters. Channel 1 corresponds to the shortest wavelengths and channel 4 to the longest. Two diffraction grating wheel assemblies (DGAs) rotate the dichroics and diffraction gratings in the optical path to select the wavelength coverage within all four channels simultaneously, dividing each channel into three spectral sub-bands: SHORT (or A), MEDIUM (or B), and LONG (or C). Figure~\ref{fig:mrs_aux} summarizes the key parameters of the MRS. To obtain a spectrum for the whole $4.9-28.8$ $\mu m$ wavelength range, the observer needs to stitch together 12 spectral bands, obtained in three exposures. In Exposure 1 both DGA positions are set to SHORT and, thus. the MRS measures the bands 1A-2A-3A-4A. In Exposure 2, the DGA positions are set to MEDIUM, thus measuring bands 1B-2B-3B-4B. Finally in Exposure 3, the DGA positions are set to LONG, thus measuring the bands 1C-2C-3C-4C. Each channel has its own IFU. Each IFU contains an image slicer that slices the imaged MRS FOV of each band into a distinct number of slices. These slices are used as entrance slits for the spectrometer and each of them is imaged as a dispersed stripe on one half of the MRS detector arrays. The MRS uses two detectors. The short-wavelength (SW) MRS detector is used in the $4.9-11.8$ $\mu m$ wavelength range (channels 1 and 2). The long-wavelength (LW) MRS detector is used in the $11.5-28.8$ $\mu m$ wavelength range (channels 3 and 4). Each of the channels have a different sampling of the sky, so the imaged FOV, sky-projected angles, and slice widths are different. Figure~\ref{fig:allbands_detimg} shows two detector images of a spatially uniform source observed with the MRS on-ground, covering all four spectral channels.
   
   \begin{figure*}
   \centering
   \includegraphics[height=10.4cm]{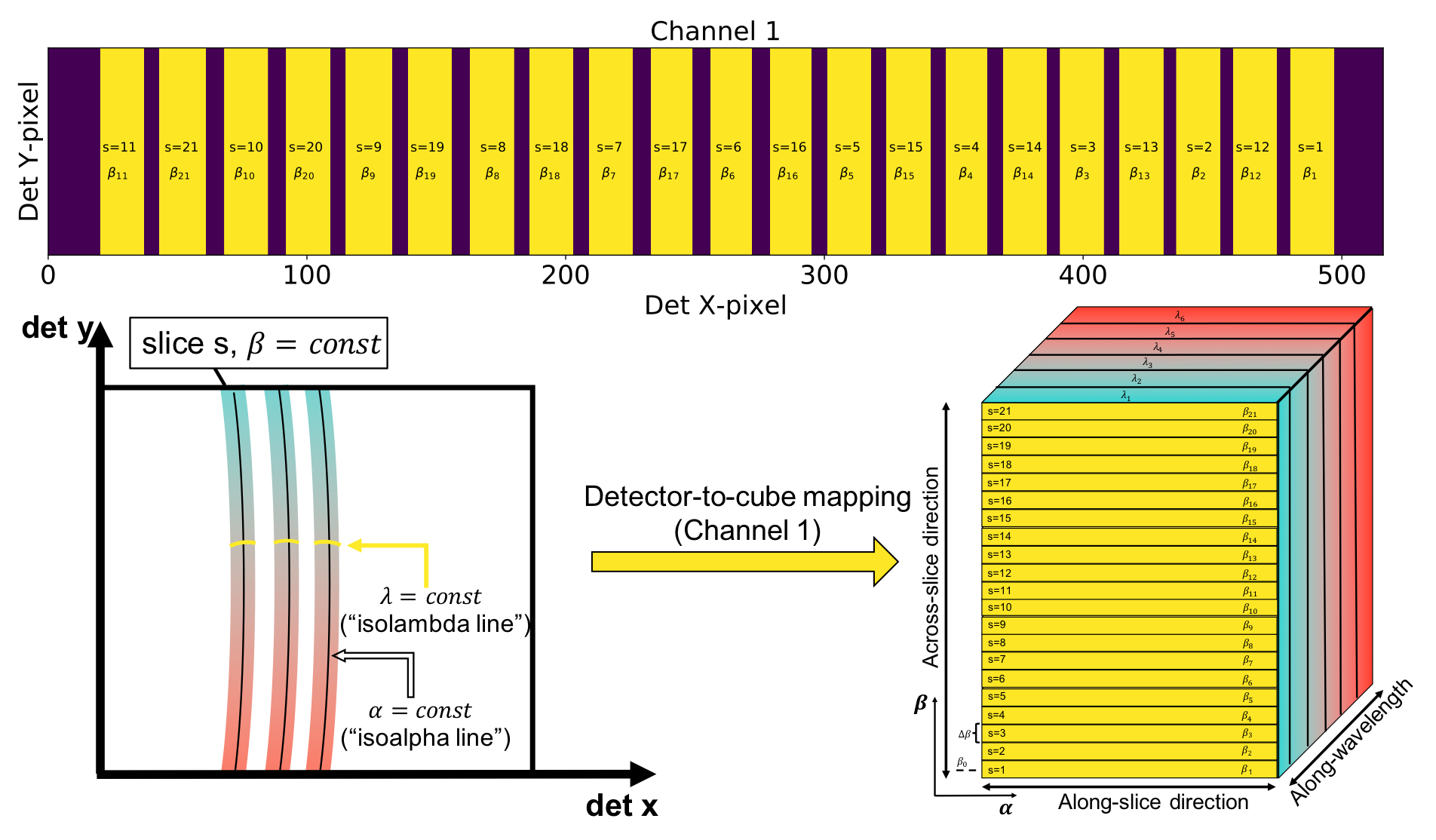}
   \caption[]
   { \label{fig:d2c}
   Relation of the detector coordinate system \textit{x/y/s} and the local MRS coordinates $\alpha$, $\beta$, $\lambda$. The transform between the detector array and the local MRS coordinates is described by the  detector-to-cube information.}
   \end{figure*}
  
  The FOV values presented in Fig.~\ref{fig:mrs_aux} are provided as-measured in local MRS coordinates. The readout of an MRS pixel can be attributed to a coordinate in different meaningful coordinate systems of the observation parameter space:
  \begin{itemize}
      \item the detector coordinate system \textit{x/y} and slice number \textit{s}: Each detector array has 1032 pixels in the horizontal direction (only 1024 of those are photosensitive) and 1024 in the vertical direction, as shown in Fig.~\ref{fig:allbands_detimg}. The different IFU image slicer slices appear as stripes on the detector and they can be numbered using a discrete integer \textit{s} with increments following the order in which the slices map on the sky. This is shown in the top plot of Fig.~\ref{fig:d2c}. Importantly, neighboring slices on the sky are not neighboring on the detector. As an example, Fig.~\ref{fig:d2c} shows how the slices are distributed on the detector in channel 1.
      \item the local MRS coordinates $\alpha$, $\beta$, $\lambda$: Each MRS channel (1, 2, 3, 4) has its own field on the sky. As such, the mapping from the detector space to the spectral cube space happens in a local frame associated to each channel. The $\alpha$-axis is defined parallel to the IFU image slicer along-slice direction, the $\beta$-axis is defined perpendicular to the $\alpha$-axis and describes the image slicer across-slice direction, and $\lambda$ represents the wavelength. An illustration of the channel 1 local MRS coordinates is shown on the bottom right part of Fig.~\ref{fig:d2c}.
      \item the global JWST telescope coordinates V1/V2/V3: V1 is the symmetry axis of the JWST telescope, V3 points towards the foldable secondary Mirror Support Structure strut, and V2 forms a right handed system for the V1/V2/V3 coordinate system. We do not use this coordinate system in this paper.
  \end{itemize}
  
  The coordinate transformations between the detector arrays, the MRS local fields, and the JWST FOV are performed using transformation matrices calibrated using on-ground data. The transform between the detector array and the channel cube is described by the detector-to-cube information. This is illustrated in the lower part of Fig.~\ref{fig:d2c}. A two-dimensional polynomial description is used for the $\alpha$ and $\lambda$ coordinates as functions of the location on the detector. These polynomials are given for each slice, \textit{s,} individually. Each detector pixel belonging to a slice, \textit{s,} has a different $\alpha$ and $\lambda$ coordinate. The polynomial description can be used to trace lines of constant $\alpha$ position ("isoalpha lines") on the detector, as well as lines of constant wavelength ("isolambda lines"). 
  
  Due to distortions introduced by the MRS optics, slices on the detector are curved. As a result, any isoalpha line and any isolambda line within a slice will also be curved. The $\beta$ coordinate is discrete and is provided through the slice number directly by $\beta(s) = \beta_0 + (s-1) \cdot \Delta\beta$. The parameter $\beta_0$ is the $\beta$ coordinate of the center of $s=1$ and $\Delta\beta$ is the slice width. The value of $\beta_0$ and $\Delta\beta$ is constant for each MRS channel and, thus, each slice \textit{s} has a unique $\beta$ value associated with it. While the size of all spaxels (spatial pixel) in the across-slice direction is fixed (equal to $\Delta\beta$), the size of a cube spaxel in the along-slice direction is defined by the along-slice spatial resolution of the MRS in a given spectral band. The size of a wavelength bin or step on the cube is defined by the dispersion of the MRS in the same spectral band.

   \subsection{Fringing in the MIRI MRS detectors}
   The MIRI instrument focal plane arrays are three arsenic-doped silicon impurity band conduction (IBC) detector arrays, also known as Si:As IBC devices \citep{love2005,miri_det_performance,miri_pasp_7}. One detector array is used for MIRI imaging, coronagraphy, and low-resolution spectroscopy. Two detectors are used for MIRI medium-resolution spectroscopy. All three MIRI detectors are operated at a temperature of 7~K.
   
\begin{figure}[t]
   \centering
   \includegraphics[height=9.5cm]{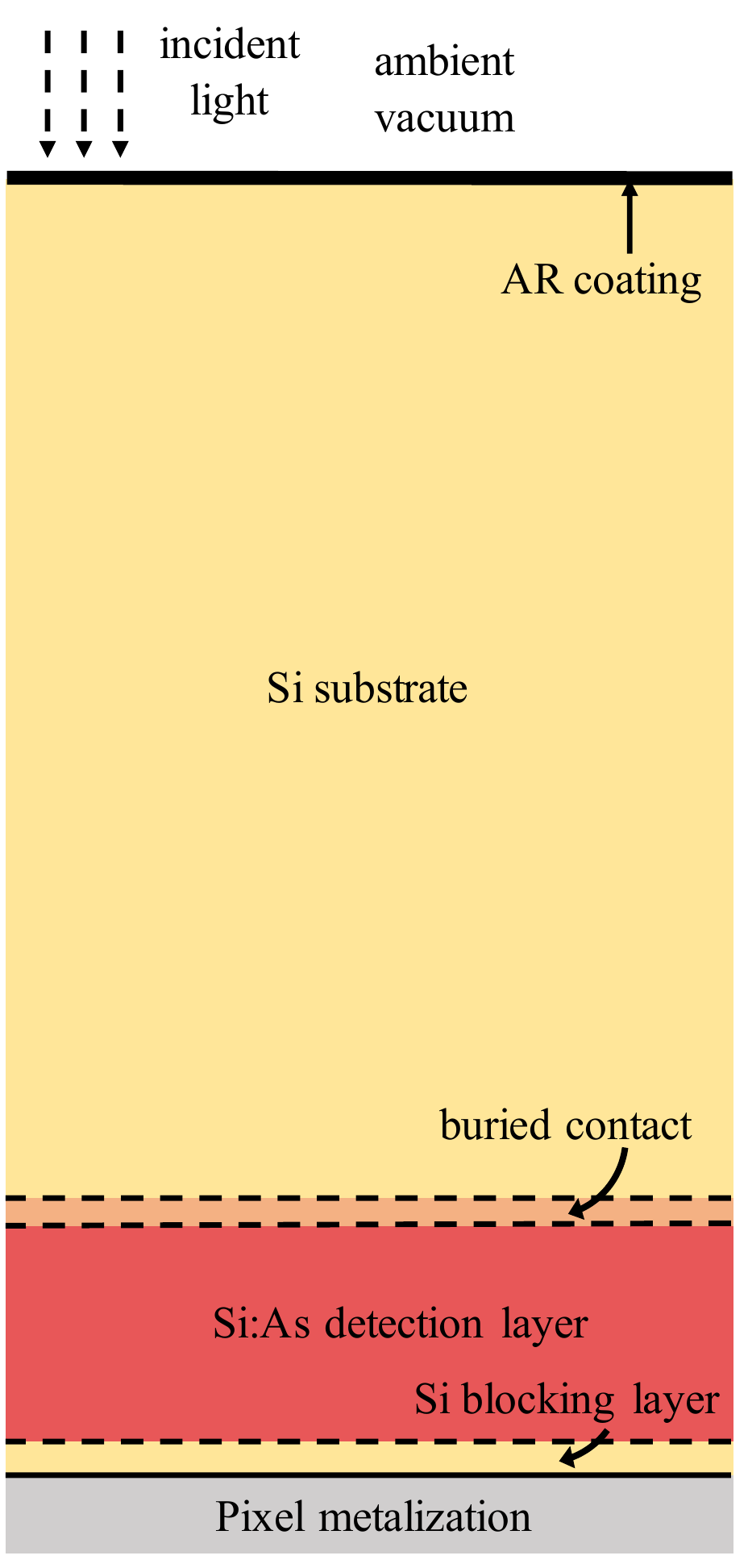}
   \caption[]
   { \label{fig:detector_layout}
   Schematic depiction of the MIRI IBC detector structure. Relative thicknesses of the layers are not to scale.}
   \end{figure}

  The architecture of the backside-illuminated Si:As IBC detector arrays can be found in \citet{love2005}. A schematic depiction of a cross-section of the array is shown in Fig.~\ref{fig:detector_layout}. Incident light travels through the anti-reflection (AR) coating. The light then travels through the high purity silicon substrate and through the ion-implanted back contact to reach the infrared-active layer. The infrared-active layer is doped with arsenic to absorb the incoming photons. An electric field is maintained across the active layer, which causes the photoelectrons to migrate to the front of the detector, where they are recorded by the MIRI read-out electronics.

   Every layer in the Si:As IBC detector arrays shown in Fig.~\ref{fig:detector_layout} has refractive properties that depend on wavelength and temperature. The temperature of the detectors is controlled with an extremely high level of accuracy, hence, we assume it remains constant during observations. Therefore, the refractive properties of the detectors depend mainly on the wavelength of the incoming light. The index of refraction and extinction coefficient of the detector layers affect how light propagates within the detectors as a function of wavelength. Coherent interference from reflections between refractive interfaces gives rise to the phenomenon called "fringing" (the detector inner layers operate as efficient Fabry-Pérot etalons). Those internal reflections also impact the quality of the MIRI imager PSF, which exhibits a so-called “cross-artifact,” which is a cross along the pixel columns and rows centered on the sources. This artifact is a result of light scattering between the detector and the readout multiplexer, which diffracts off the pixel grid periodic lattice (Gáspár et al., in prep).
   
   The amplitude, frequency, and phase of the fringes depend on the refractive and geometric properties of the detector. Fringes in the MIRI MRS can be discerned in Fig.~\ref{fig:allbands_detimg} as the periodic bright and dark modulation of the light in the dispersion direction. 
   
   Fringes can impact the scientific exploitation of the data in a number of ways:
   \begin{itemize}
       \item Fringe modulation in the extracted spectra hides weak spectral lines.
       \item Fringes introduce larger errors on the determination of the spectral continuum.
       \item The shape of spectral lines is distorted, regardless of whether these are in absorption or emission.
       \item A larger uncertainty is imposed on the spectral line-to-continuum ratio.
       \item Fringes introduce a modulation of the apparent centroid of a point source on the detector.
       \item Reconstructing 3D spectral cubes using pixels from different parts of the detector introduces correlated noise. This is important, for instance, when attempting to perform high-contrast imaging with the MRS.
   \end{itemize}
   
   \subsection{Commonly used fringe correction methods and their limitations}
   \label{subsec:limitations}
    
    There are two common fringe correction methods used to remove fringing from detector images. The first fringe correction method consists of calibrating the data with the use of a "fringe flat." A fringe flat can either be measured on the ground, or using an onboard calibration source measured right before or after a scientific observation. The data are then divided by the fringe flat, and ideally the fringes disappear from the scientific image. 
    
    The second method, often applied after division by a fringe flat, is an empirical fringe correction method. The (residual) fringes are first modeled using a series of sinusoids and then the data are divided by the model fringe pattern. Such a method was used to correct the fringes in the ISO SWS, Spitzer IRS, and the Herschel HIFI data \citep{sws_fringes_and_models,fringes_sirtf_irs,herschel_hifi}.
    
    The above two methods have the following limitations. Firstly, as demonstrated below, sources of different spatial profiles yield different fringe patterns on the detector. Applying a fringe flat derived from an extended source to a point source or a source with spatial structure effectively introduces systematic and correlated noise between spectral bins. Secondly, empirical fringe correction methods attempt to disentangle the fringe signature from the real source spectral signature. Spectral features that show periodicities that are commensurate with the fringe period may erroneously be interpreted as a fringe signature. This has implications for science as the correction should of course neither change nor, in the worst case, remove the signature of spectral features. Thirdly, if the data have a low signal-to-noise ratio, the empirical correction is bound to yield large uncertainties.
    
    In the case of MIRI, the planned correction method is to apply a division by a fringe flat, followed by an empirical (“residual”) fringe correction \citep{miri_mrs_cal_pipeline}. A fringe flat has been measured on the ground for each spectral band of the MRS using an extended, spatially uniform blackbody source. In practice, perfectly flat and uniform extended sources don’t exist in the sky. MIRI will have to deal either with point sources (e.g., exoplanet hosts), or with semi-extended sources with spatial structure (disks, galaxies). As such, it is important that the applied fringe correction be able to accurately reduce the impact of fringing on MRS spectra for sources of different spatial structures. This would also reduce or even omit the need for an empirical residual fringe correction, avoiding the intrinsic limitations linked to such a method.

\section{Data and methodology}
\label{sec:spectral_extraction}
\subsection{MRS point source data}
    Two CV test campaigns were held at NASA-GSFC between 2014 and 2016. The transmission of the optical elements used during the CV tests yielded useful signal for the fringe analysis only up to 6 $\mu m$. Therefore, this fringe analysis is limited to MRS band 1A. Extending the analysis to the other MRS bands would require observing a point source with a (preferably) bright spectral continuum from 4.8~$\mu m$ to 28.8~$\mu m$. Such sources will be observed post-launch during the MIRI instrument commissioning phase. From the CV point source measurements in band 1A, all pointings (32 in total) were used in the analysis, covering the entire MRS FOV. 
    
    The point source used during CV testing was a IATF4 broad-band infrared light-emitting diode (LED) purposed for the measuring of the MRS PSF. The manufacturer of the point source is Ioffe LED, Ltd. (based in St Petersburg, Russia); the LED is based on the Ioffe OPLED70 model \citep{ioffeLED}. The PSF is defined by a pinhole plate mounted in front of the source. A sapphire diffuser placed in the source optical path affects the intrinsic spectral energy distribution of the source \citep{surmet}. The resulting source spectral energy distribution peaks at 5.4~$\mu m$ (peak integrated flux of approximately 5 Jansky) and rapidly falls off at 6.3~$\mu m$. Exposures were taken at each pointing position with the LED source turned off for background subtraction. The resulting PSF has a full width at half maximum (FWHM) of 0.317 arcseconds in along-slice ($\alpha$) direction and a FWHM of 0.226 arcseconds in across-slice ($\beta$) direction.
   
    \begin{figure}[t]
    \centering
    \includegraphics[height=7.5cm]{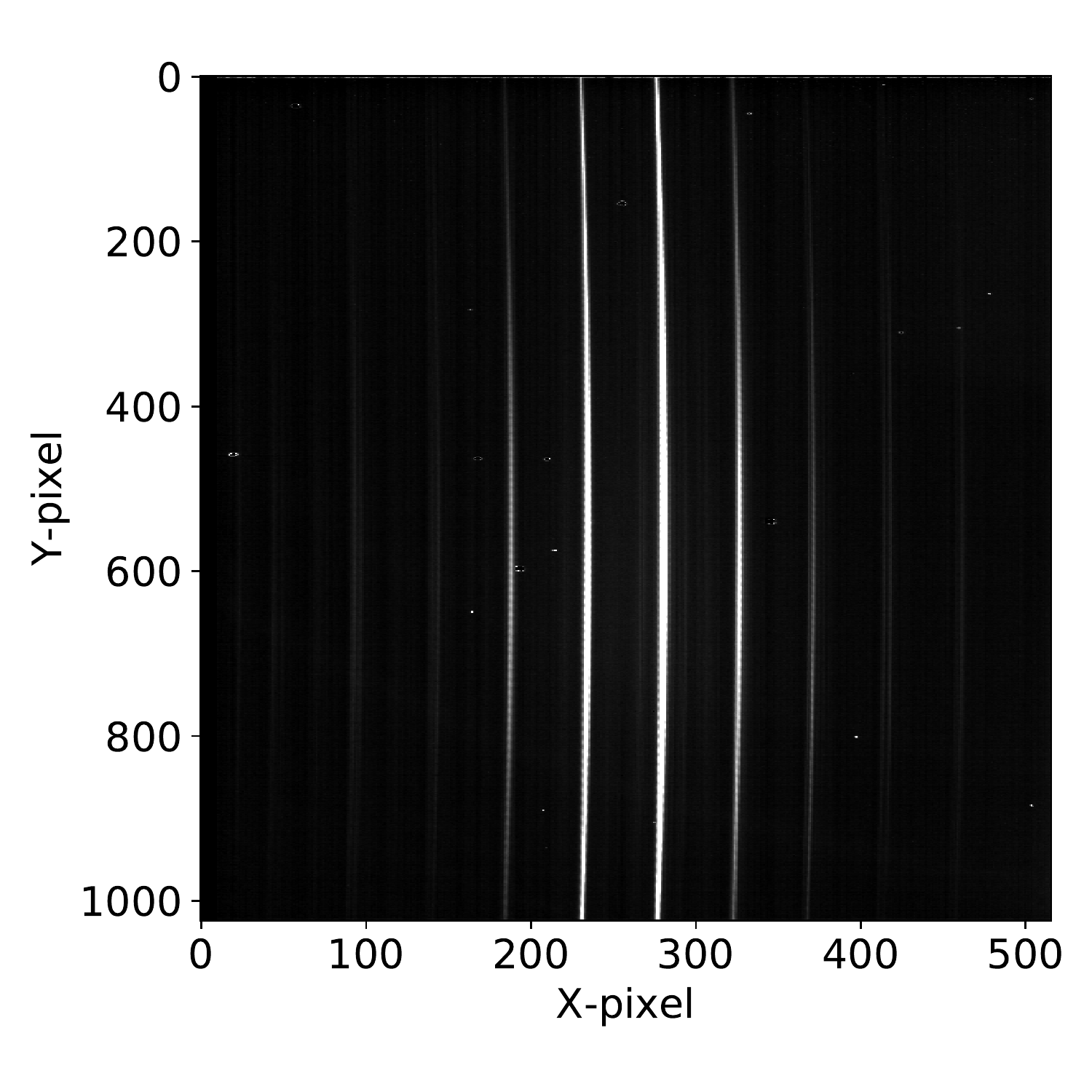}
    \caption[example]
    { \label{fig:point_source_det_img}
    MRS point source spectrum on detector plane image. The intensity in the different traces varies as a function of the given part of the PSF that is sampled (core versus wings). The signal contrast has been selected to allow for the visualisation of the slices in which the point source is significantly detected.}
    \end{figure}
  
\subsection{Extracting MRS spectra from detector plane images}
   In Fig.~\ref{fig:point_source_det_img}, we show a typical MRS detector image of a point-like source observed during the CV campaign. Because neighboring IFU slices on the sky are not neighboring on the detector (as shown in the top plot of Fig.~\ref{fig:d2c}), the source point-spread-function is split into a distinct number of traces. We define a "trace" as the collection of isoalpha lines that contain the bulk of the PSF in a given slice. The intensity in the different traces shown in Fig.~\ref{fig:point_source_det_img} varies as a function of the given part of the PSF that is sampled (core versus wings).
   
\begin{figure}
   \centering
   \includegraphics[height=3.3cm]{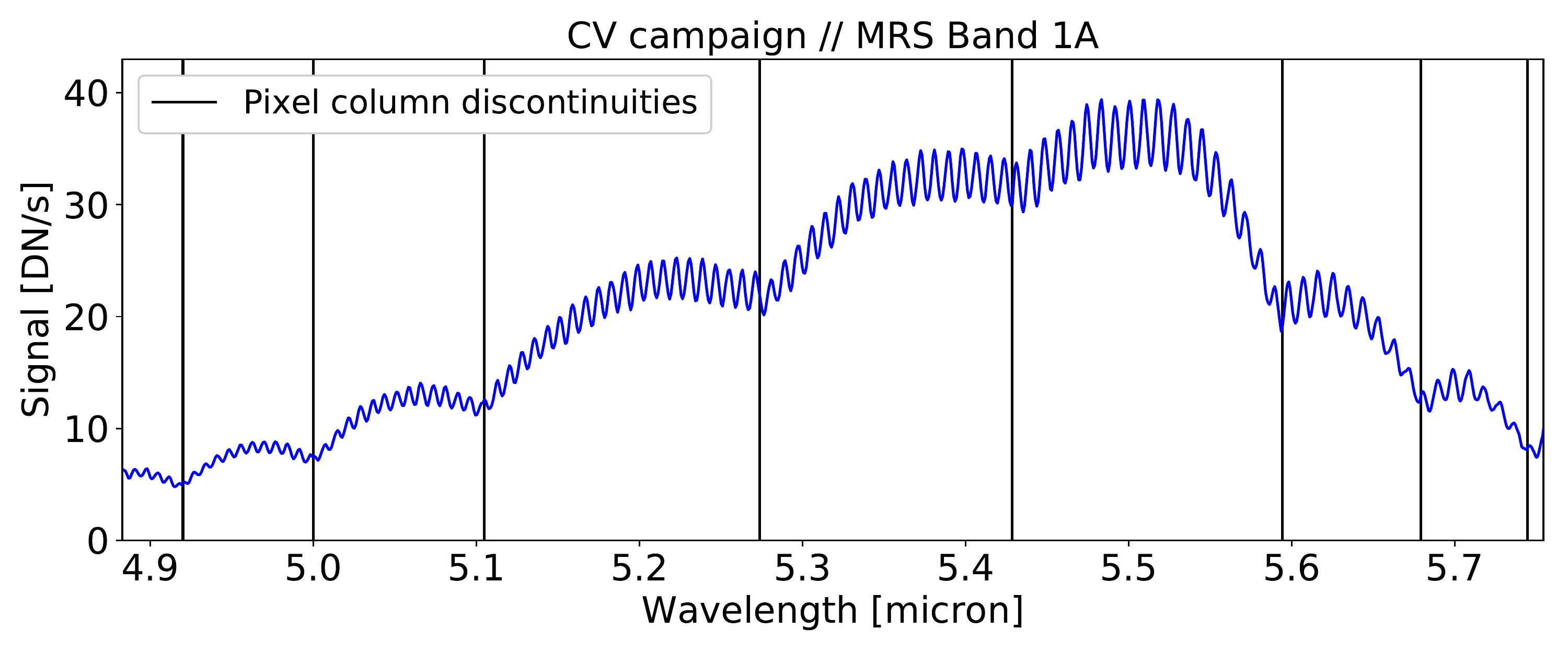}
   \caption[example]
   { \label{fig:point_source_spectrum}
   Point source spectrum extracted from an isoalpha line on an MRS detector plane image (acquired during the JWST ISIM CV testing). The signal units are given in Digital Number (DN) per second. The pixel column discontinuities occur at the rows where the curved isoalpha line is "crossing" from one column to a neighboring one.}
   \end{figure}

   Every MRS detector pixel has a corresponding triplet of coordinates in the local MRS coordinates ($\alpha$, $\beta$, $\lambda$). The simplest way to extract a spectrum is by selecting a slice,~\textit{s,} corresponding to a unique $\beta$ value and identifying an isoalpha line in that slice. An isoalpha line is identified by first defining a desired spatial position in $\alpha$ (for instance $\alpha$ = 0 arcseconds). Afterwards, for each detector row the pixel in slice
  ~\textit{s} with the $\alpha$ coordinate closest to the defined location is determined. By plotting the flux value of each of these 1024 pixels (1024 rows on detector), a spectrum is effectively extracted. Attached to that spectrum, the wavelength coordinate of each pixel is used to uniquely define each pixel in the spectral direction. An example result of this procedure is shown in Fig.~\ref{fig:point_source_spectrum}, where the slice with the largest integrated intensity on the detector is used. The signal units are given in digital number (DN) per second.
   
   In Fig.~\ref{fig:point_source_spectrum}, we denote, using vertical black lines, the discontinuities in the X-pixel position on the detector. The discontinuities occur at the rows where the curved isoalpha line is "crossing" from one column to a neighbouring one. We note the arcing of the spectral baseline between pixel discontinuities; between two discontinuities the signal increases, reaches a maximum at the middle, and then decreases. This variation in signal is caused by the sampling of the PSF by the pixels in the isoalpha line. 
   
\begin{figure}[h]
   \centering
   \includegraphics[height=3.5cm]{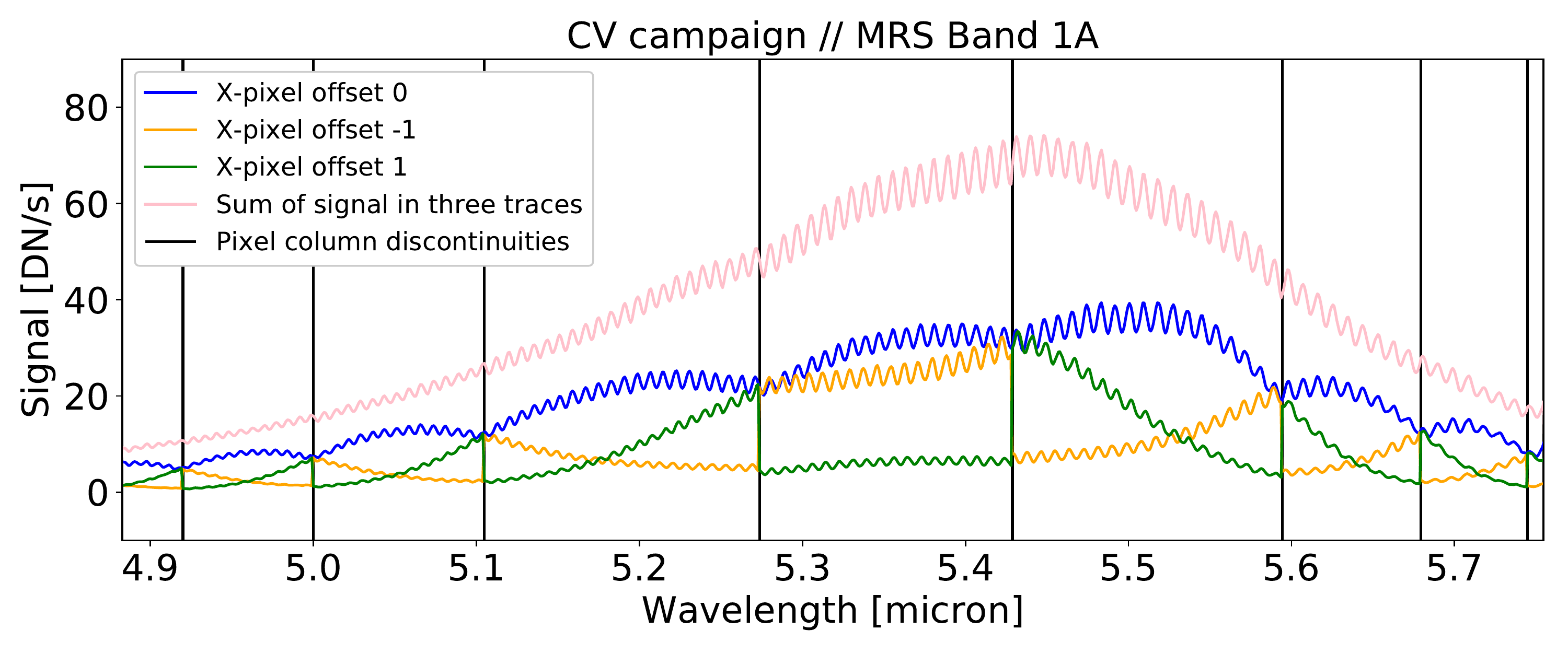}
   \caption[example]
   { \label{fig:summed_spectrum}
   Point source spectra extracted from three isoalpha lines on the detector. A point source integrated spectrum is computed by summing the signal in the three isoalpha lines. The resulting spectrum represents the LED source spectrum multiplied by the optical transmission from the LED to the MIRI instrument, multiplied by the MRS instrument response in band 1A (approximately 1.2 (DN/s) / milliJansky).}
   \end{figure}

   To get the full integrated spectrum of a point source, all the parts of the PSF need to be accounted for. A partial fulfillment of this requirement is shown in Fig.~\ref{fig:summed_spectrum}. Here, we plot three neighboring isoalpha lines as well as their summed signal in the spectral direction. The three isoalpha lines are defined as follows:
   
   \begin{itemize}
   \item the central isoalpha line (X-pixel offset 0) approximately follows the point source centroid (peak of the PSF). This is the spectrum of Fig.~\ref{fig:point_source_spectrum}.
   \item The second isoalpha line (X-pixel offset -1) is fully offset from the “central” isoalpha by one horizontal pixel to the left on the detector. This line follows the PSF left wing.
   \item The third isoalpha line (X-pixel offset 1) is fully offset from the “central” isoalpha by one horizontal pixel to the right on the detector. This trace follows the PSF right wing.
   \end{itemize}
   
   If the signal of the three isoalpha lines is summed in the spectral direction, an integrated point source spectrum emerges. This spectrum is the LED source spectrum multiplied by the optical transmission from the LED to the MIRI instrument, multiplied by the instrument response in band 1A (approximately 1.2 (DN/s) / milliJansky). Rather than investigating the fringes from the integrated source spectrum, in our study the fringes in each isoalpha line are investigated separately. This approach allows us to link the pixel position with systematic changes in the spectral modulation of the fringes.

\subsection{Extracting the fringe information from MRS spectra}
\begin{figure}[h]
   \centering
   \includegraphics[height=5.9cm]{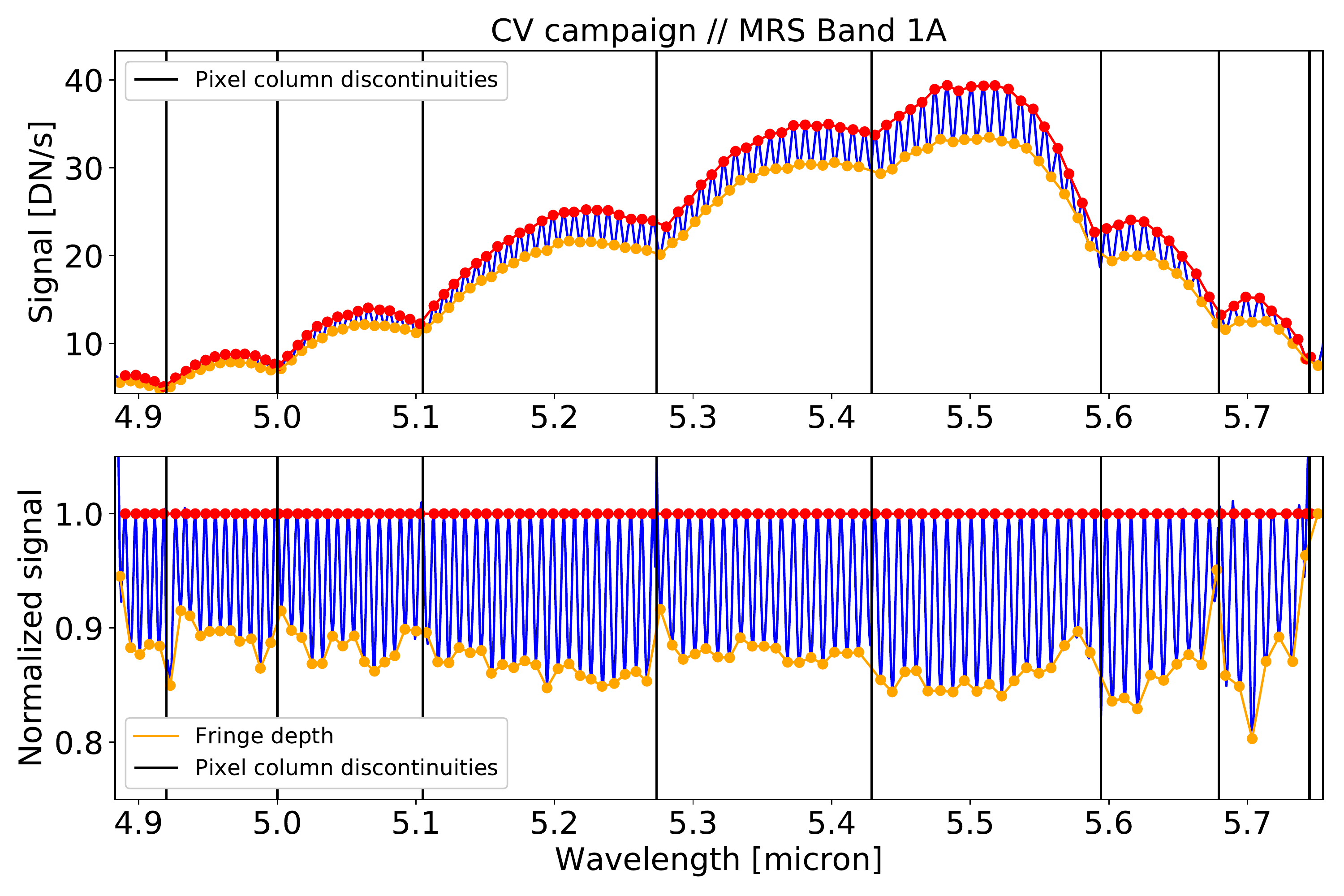}
   \caption[example]
   {\label{fig:troughs_fringe}
   Top plot: Point source spectrum in single isoalpha line. Bottom plot: Point source spectrum normalized to the fringe-peak continuum. The fringe depth is also delineated.}
   \end{figure}

   A fringe transmission spectrum can be derived from the signal in each isoalpha line. To do this, first we identify the fringe peaks (local maxima) in the sampled signal. Secondly, an interpolating cubic spline is fitted using the identified fringe peaks as knots. The spline fitting is done in each spectrum segment bound between two pixel column discontinuities. The result of this procedure defines the fringe-peak spectral continuum. Thirdly, to get the detector fringe transmission, the spectrum is normalized to the fringe-peak spectral continuum. The top plot of Fig.~\ref{fig:troughs_fringe} shows the fringe peaks that define the spectral continuum. Similarly the fringe troughs are also identified. The latter can be used in the normalized spectrum to establish the fringe depth as a function of wavelength. The bottom plot of Fig.~\ref{fig:troughs_fringe} shows the normalized spectrum. The fringe depth is illustrated by the linear order spline that is fitted through the fringe troughs.

\begin{figure}[t]
   \centering
   \includegraphics[height=3cm]{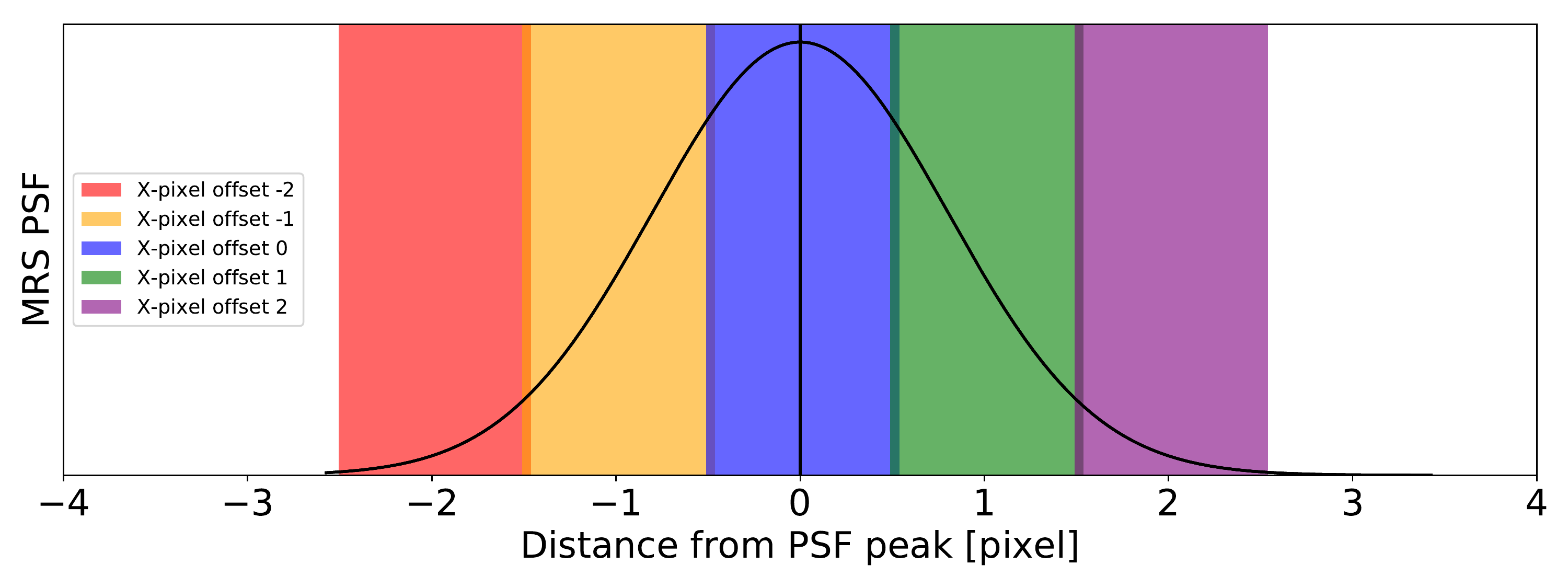}
   \caption[example]
   { \label{fig:psf_isoalpha_coverage}
   Coverage of MRS PSF in neighboring isoalpha lines.}
   \end{figure}

   The transmission spectrum shown in the bottom plot of Fig.~\ref{fig:troughs_fringe} is derived by considering a specific portion of the MRS PSF near the peak. At every sample of the examined isoalpha line, a different part of the PSF is sampled. We show this in Fig.~\ref{fig:psf_isoalpha_coverage}, where the MRS PSF is approximated with a Gaussian distribution of corresponding spatial size to the real MRS PSF. We also show the spatial coverage of four additional isoalpha lines (the dominant part of the MRS PSF in band 1A is covered by approximately five isoalpha lines). 
   
\begin{figure}[t]
    \centering
    \includegraphics[height=8.cm]{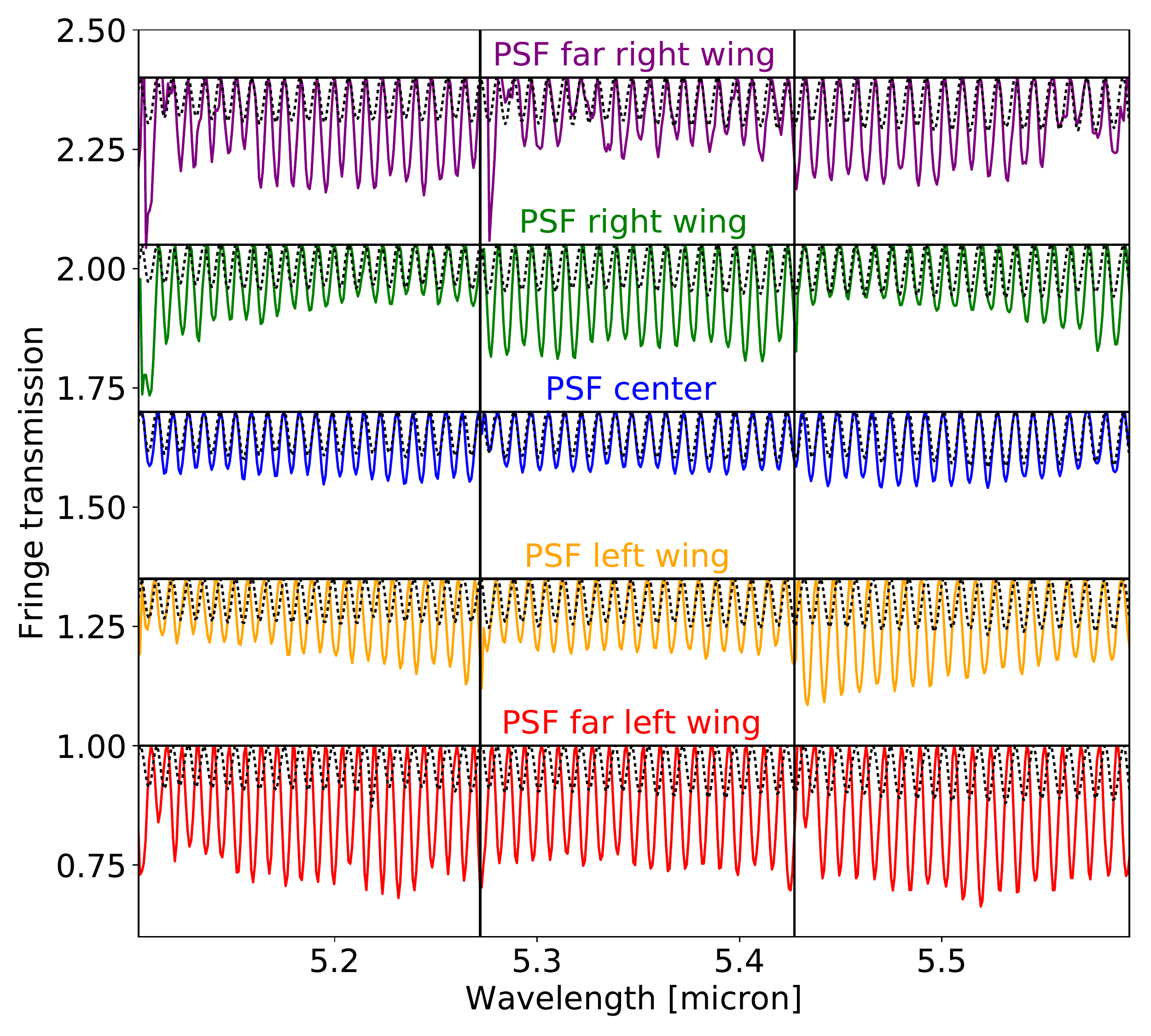}
    \caption[example]
    { \label{fig:ext_vs_point_source}
    Point source fringe transmission computed from five neighboring isoalpha lines (colored curves offset on the vertical axis by 0.35). The fringe transmission of the extended source observed at the same location is overplotted (dotted black curves). The extended source fringe amplitude, frequency, and phase is the same at all locations of the PSF. A reduced wavelength range is shown for better visibility.}
    \end{figure}

   In Fig.~\ref{fig:ext_vs_point_source}, we show the fringe transmission derived from the same five isoalpha lines used in Fig.~\ref{fig:psf_isoalpha_coverage}. The fringe transmission of the extended source used to derive the MRS fringe flat is overplotted with a dotted black line. Interestingly, the extended source fringe transmission matches the point source fringe transmission best at the PSF center, where the flux density peaks, that is to say that the fringe phase matches and there is but a small discrepancy in the fringe depth. We attribute this to the fact that the pixel illumination is most similar between the extended source and the point source at the PSF peak (spatially uniform illumination), as illustrated in Fig.~\ref{fig:psf_isoalpha_coverage}. When other portions of the PSF wings are sampled, the situation starts to change. In the left wing of the PSF,  we see that the fringe phase does not match everywhere and there are larger differences in fringe depth. This gets worse at the far-left wing of the PSF. However, looking at the far right wing of the PSF, the situation is not mirrored. The fringe phase of the extended source is a much better match, however, the fringe depth is still discrepant.
   
   The fringe depth changes drastically from one spectrum to another, and significant jumps are recorded past pixel column discontinuities. The geometric and refractive properties of the detector do not change fast enough over the detector and over the examined wavelength range to explain this change. It could be presumed that the fringe depth is dependent on the absolute signal level of the spectra. This is, however, not the case. Tests performed on blackbody observations of different effective temperatures (400~K, 600~K, and 800~K) have shown the same fringe transmission despite the widely different absolute signal levels (factors of 1, 10, and 100 higher relative flux). As such, there is something else that is causing this  difference in contrast  and we explore this issue in Sect.~\ref{sec:fringe_amplitude}.

\section{Results and discussion}
\subsection{Understanding the point source fringe contrast}
\label{sec:fringe_amplitude}
\begin{figure}[h]
    \centering
    \includegraphics[height=8cm]{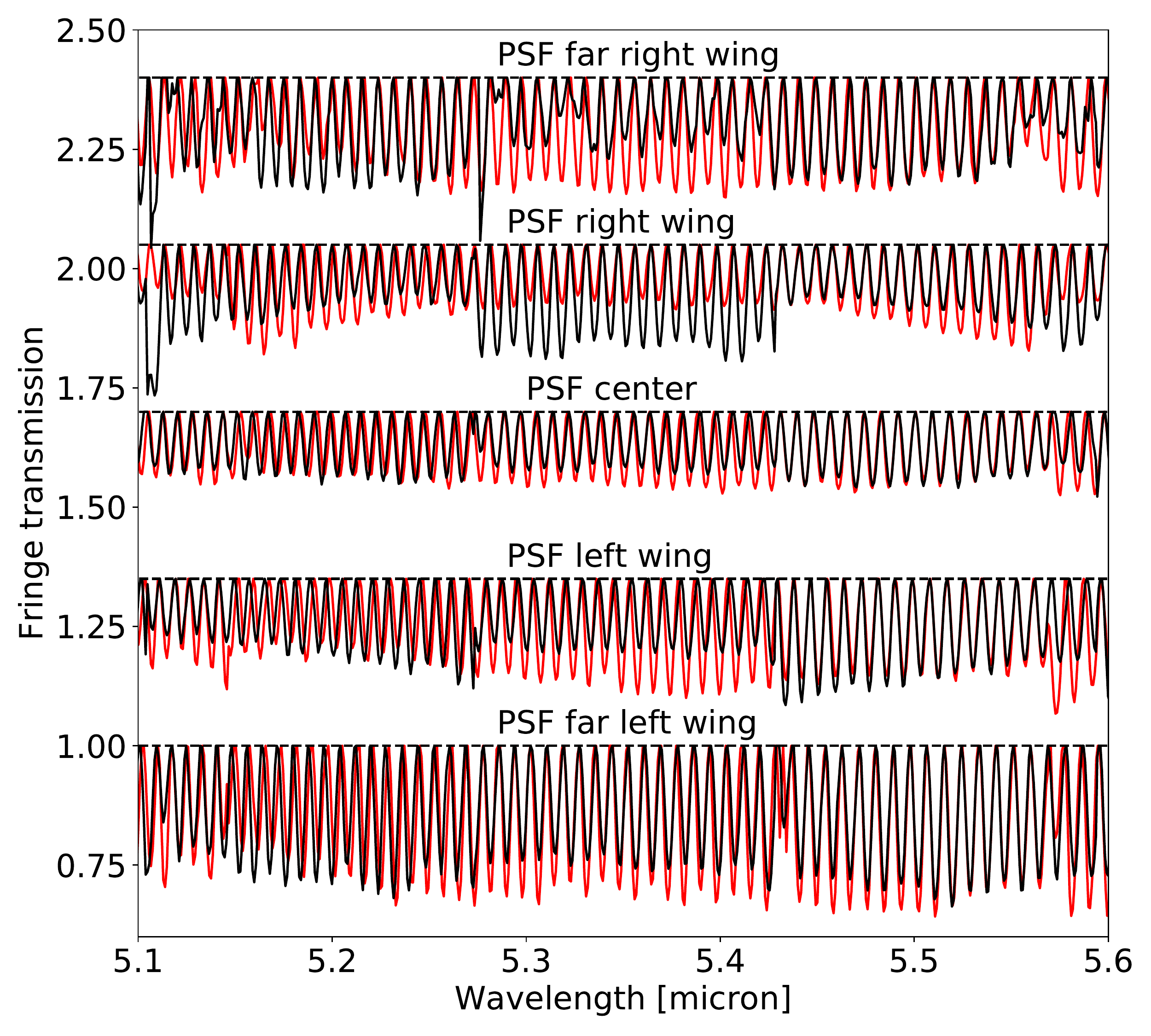}
    \caption[example]
    { \label{fig:point_source_P0_vs_P1}
    Comparison between the fringe transmission derived from the same point source positioned at two different locations in the MRS FOV, namely, two different locations on the MRS detector. The observations were acquired in MRS band 1A.}
    \end{figure}

   In Fig.~\ref{fig:point_source_P0_vs_P1}, we compare the fringe transmission derived from the same point source at two different locations in the MRS FOV, thus, at two different locations on the MRS detector. At first sight, the fringe depth and phase appear erratic. In this section, we aim to demonstrate that the discrepancy in fringe depth does not depend on the point source position in the MRS FOV but, rather, on the part of the PSF that is sampled by any given pixel. To do so, we analyze all available point source ground-based data taken at several pointings and link them in terms of what part of the PSF is being sampled at every point.
   
\begin{figure}[h]
  \centering
  \includegraphics[height=4.5cm]{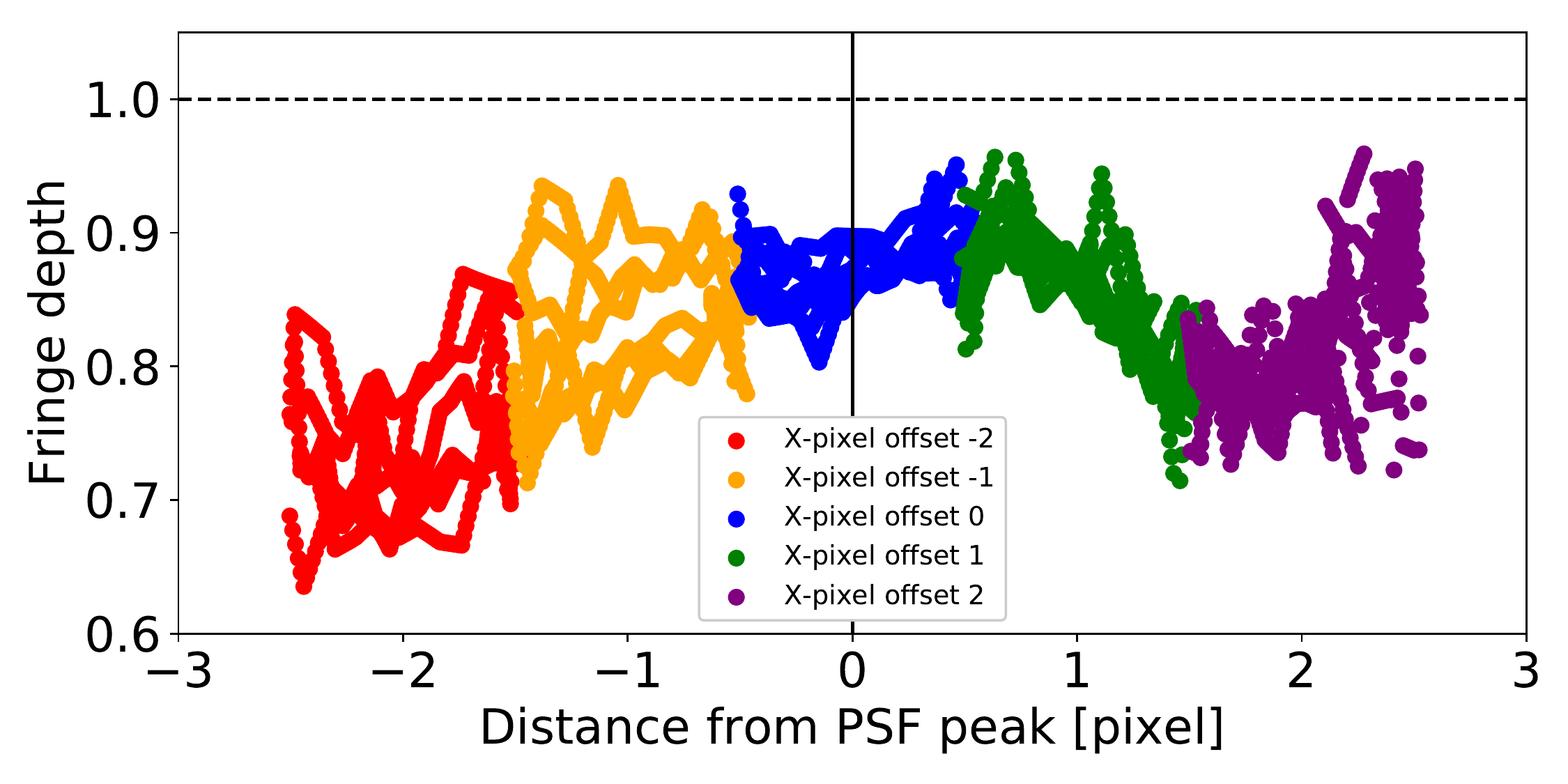}
  \caption[example]
  { \label{fig:psf_fringe_amplitude}
  Fringe depth as a function of distance from the PSF peak.}
  \end{figure}

   We want to compare the fringe depth at every sample of every isoalpha line with the distance of a said sample from the observed PSF peak. We define the "distance from the PSF peak" as the $\alpha$-coordinate of a given sample (in arcseconds) divided by the MRS plate scale at that position (in units of arcseconds/pixel) minus the centroid of the observed point source (fitted PSF peak). In Fig.~\ref{fig:psf_fringe_amplitude}, the fringe depth determined from the spline connecting the fringe troughs (as shown in the bottom plot of Fig.~\ref{fig:troughs_fringe}) is plotted as a function of the distance from the PSF peak. We show this for a single dataset, namely for the five isoalpha lines shown in Fig.~\ref{fig:ext_vs_point_source}.
   
   Based on the shape of the global curve shown in Fig.~\ref{fig:psf_fringe_amplitude}. we see that, firstly, the variation in the fringe depth as a function of distance from the PSF peak varies smoothly. Secondly, the fringe contrast is least at the center and increases towards the edges. Thirdly, the data of "X-pixel offset 2" (purple points) show a prominent uptick.
   
\begin{figure}[t]
   \centering
   \includegraphics[height=8cm]{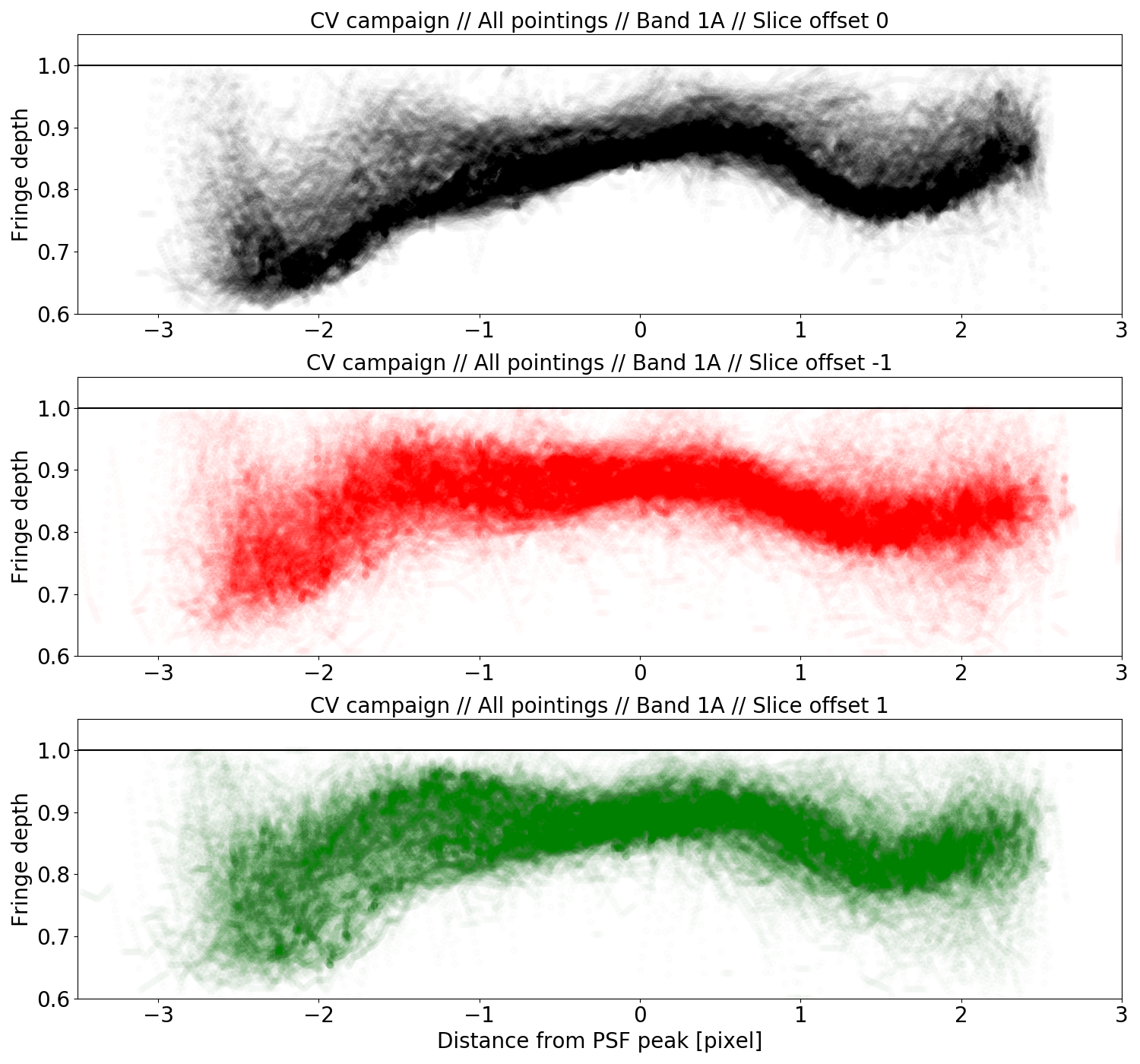}
   \caption[example]
   { \label{fig:all_points_three_slices}
   Fringe depth as a function of distance to the PSF peak. All CV pointings are used. Three slices on the detector are examined, i.e. different parts of the MRS PSF in the $\beta$-coordinate direction.}
   \end{figure}

   A single point source pointing was used to derive the trend in Fig.~\ref{fig:psf_fringe_amplitude}. Furthermore, only a single slice, \textit{s,} was used, namely, the slice with the highest intensity on the detector, containing the portion closest to the PSF peak in the $\beta$-coordinate direction. In Fig.~\ref{fig:all_points_three_slices}, all the available CV pointings are used. Three plots are shown, all pertaining to the slice~\textit{s} with the highest intensity on the detector for each respective pointing, as well as the result from slice \textit{(s-1)} and slice \textit{(s+1)}. Slice \textit{s}, \textit{(s-1)} and \textit{(s+1)} probe different parts of the PSF on the $\beta$-coordinate. Fig.~\ref{fig:all_points_three_slices} shows that the same smooth behavior is observed regardless of the location of the point source in the MRS FOV.
   
   The systematic trends in Fig.~\ref{fig:all_points_three_slices} have been established over full isoalpha lines (using all 1024 samples in a line). This means that the trends exist independently from the exact centroid of the PSF with respect to the detector pixel grid. We are, therefore, left with three main control parameters for the fringe depth within a given pixel, these being: (A) the fraction of the PSF falling within the pixel, (B) the purely wavelength-dependent reflectance of the MIRI detector layers, and (C) the change in illumination angles across the detector. 
   
\begin{figure}
   \centering
   \includegraphics[height=3.cm]{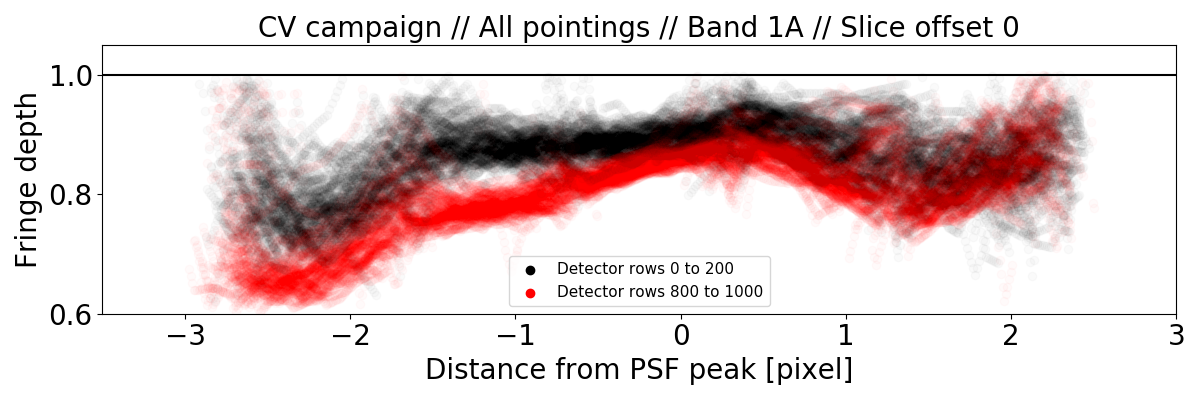}
   \caption[example]
   { \label{fig:all_points_three_slices_detdiffparts}
   Fringe depth as a function of detector position. The bulk of the dispersion in Fig.~\ref{fig:all_points_three_slices} isn't noise, but results from the spectral distribution of the fringe characteristic.}
   \end{figure}

   Point (A) is derived from the smooth and reproducible relations in figures \ref{fig:psf_fringe_amplitude} and \ref{fig:all_points_three_slices}. Point (B) is a product of the optical properties of the detector-constituting layers as a function of wavelength. Point (C) results from the different incidence angles in the spatial and spectral direction on the detector, the effect of which is visible in Fig.~\ref{fig:all_points_three_slices_detdiffparts}. Two parts of the detector are considered in Fig.~\ref{fig:all_points_three_slices_detdiffparts} (top and bottom of detector plane). If point (C) was not a control parameter, the two curves would be parallel. The rotation between the two curves is caused by the smooth variation of the incidence angles across the MRS detector dispersion direction. The incidence angle variation in the spatial direction cannot be discerned in the trend (smaller than the scatter in the trend). The large scatter at the edges of the PSF is predominantly caused by the noise in the data since, by definition, these correspond to pixels where the signal is very low.
   
\subsection{Understanding the point source fringe phase}
\label{sec:fringe_phase}
\begin{figure}[h]
   \centering
   \includegraphics[height=2.9cm]{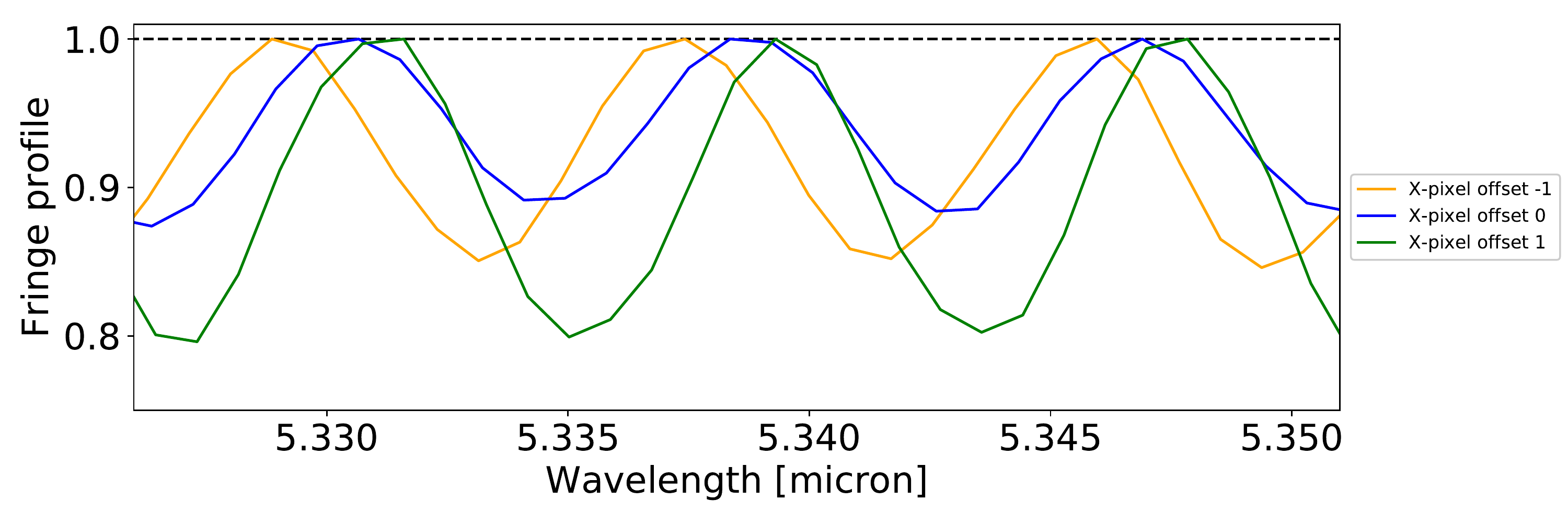}
   \caption[example]
   { \label{fig:fringe_phase}
   Fringe phase in three different isoalpha lines sampling different parts of the PSF.}
   \end{figure}

   In Fig.~\ref{fig:fringe_phase}, the central wavelength range of the fringe transmission in Fig.~\ref{fig:ext_vs_point_source} is shown for three of the five isoalpha lines. There appears to be a systematic phase-shift as a function of which portion of the MRS PSF is sampled. The peaks of the fringe profiles, offset by -1, 0 and 1 X-pixel with respect to the maximum illumination, appear in a sequential order along the wavelength axis. Offset -1 (orange) is the bluest in wavelength, offset 0 (blue) is in between and offset 1 (green) is the reddest in wavelength.
   
\begin{figure}
   \centering
   \includegraphics[height=6cm]{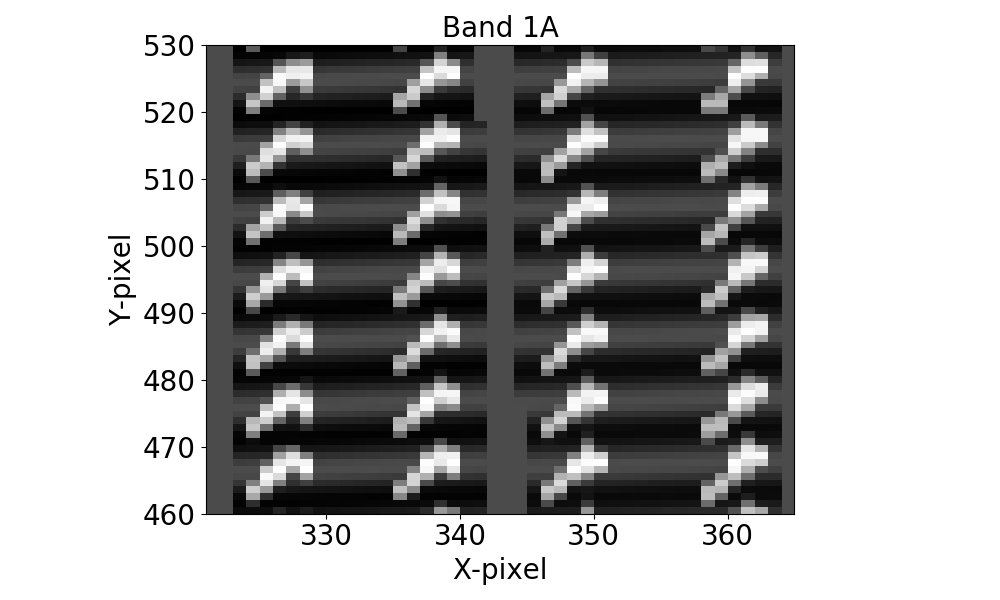}
   \caption[example]
   { \label{fig:fringe_img_CVcampaign_zoom}
   Fringe transmission profile derived from a point source observed at four different locations in the MRS FOV. The point source fringe transmission is overplotted on the extended source fringe transmission.}
   \end{figure}

   In order to analyze the systematic nature of the fringe phase-shift, the fringe transmission shown in Fig.~\ref{fig:fringe_phase} needs to be visualized in the detector space. This is shown in Fig.~\ref{fig:fringe_img_CVcampaign_zoom}. The fringe transmission for five isoalpha lines around the core of the MRS PSF (X-pixel offset -2, -1, 0, 1, 2 in slice with highest intensity) is determined and shown on the detector for CV pointings. These are overplotted on the fringe transmission of an extended source, which is used to derive the fringe flat on ground.
   
   Looking at the point source fringe transmission, a linear trend can be seen that corresponds to the trend in phase noted in Fig.~\ref{fig:fringe_phase}. Fringe peaks appear at larger and larger y-positions (higher and higher wavelengths) along the x-axis. This is true from traces -2 to trace +1, but may nevertheless be challenged between traces +1 and +2. The same trend is found at other detector locations. Figure~\ref{fig:fringe_img_CVcampaign_zoom} shows that the point sources reproduce a similar pattern (linear trend in phase) irrespective of their location in the FOV. To put it differently, there is no smooth transition between the two edges of a slice and the pattern is replicated at the different PSF locations.
   
   In order to quantify the change in the phase of the fringes as a function of the given part of the PSF that is sampled, we need to define a reference, that is, a location where the fringe phase is, by definition, zero.  For this, we decided to use the fringes of the spatially homogeneous extended source observed on-ground. The point source phase-shift systematic is then quantified based on the following three steps.

   In the first step, the peak-to-trough regions around all point source fringe peaks and extended source fringe peaks are fitted sequentially with a Fabry-Pérot transmittance function as given by Eq. \ref{eq:FPfunc} \citep{lipson1969}. This type of function mathematically describes the transmission profile of a Fabry-Pérot etalon (\textit{T$_e$}) with a coefficient of finesse \textit{F}. The optical thickness of the etalon is given by \textit{D}, where \textit{D} is proportional to the refractive index of the material between the two mirrors (\textit{n}) times the distance between the mirrors (\textit{l}). The parameter $\theta$ represents the light incidence angle onto the etalon and $\sigma$ symbolizes the incoming light wavenumber (equal to 1/$\lambda$).
       
       \begin{equation}\label{eq:FPfunc}
           T_e = \frac{1}{1 + F \cdot \sin(2\pi D \cos(\theta) \sigma)^2 } = \frac{1}{1 + F \cdot \sin(2\pi \hat{D} \sigma)^2 }
       .\end{equation}
   
\begin{figure}[h]
   \centering
   \includegraphics[height=5.4cm]{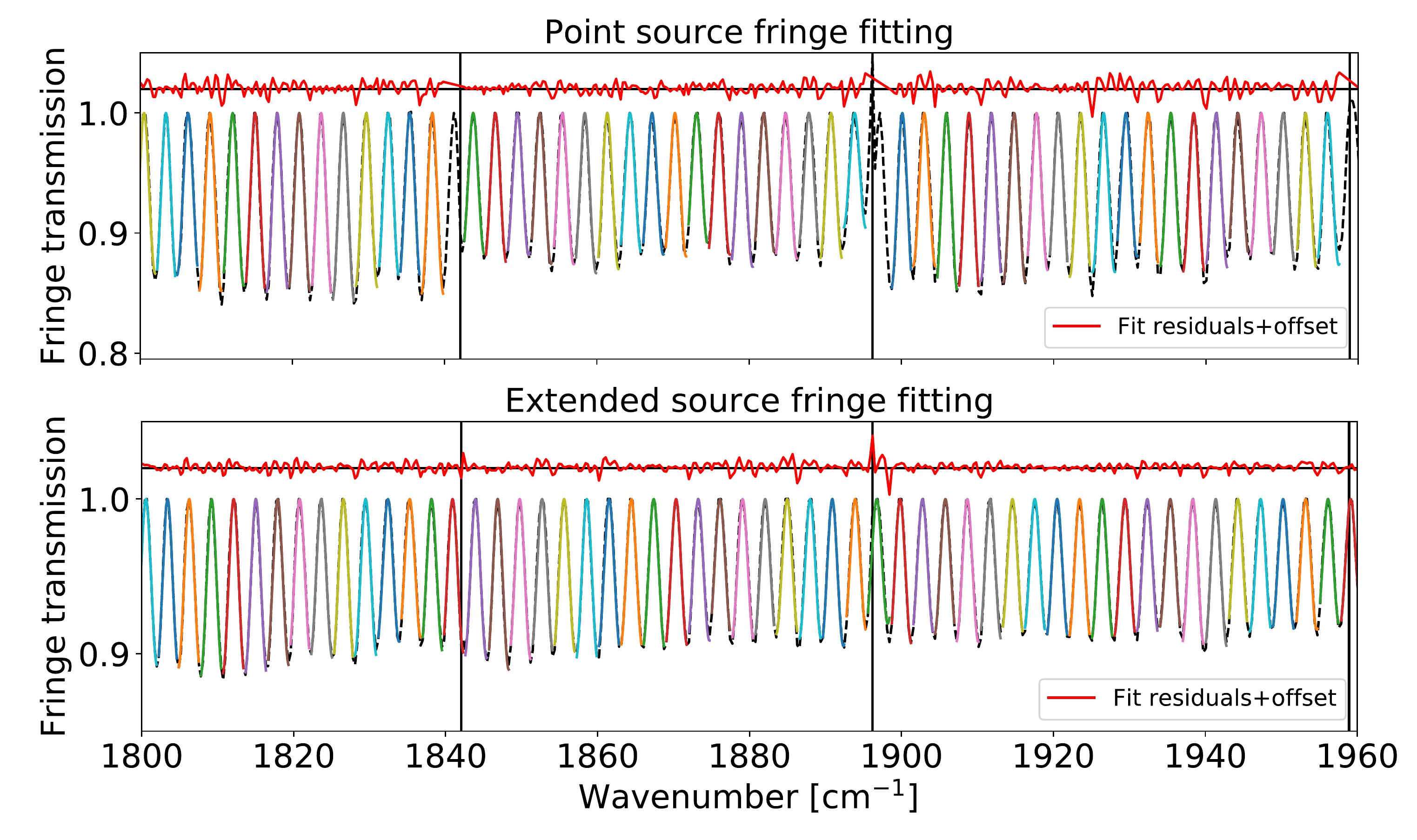}
   \caption[example]
   { \label{fig:fringe_fitting}
   Fitting of point source and extended source fringe peaks over a subset of the wavelength range. The fits are overplotted on the original fringe transmission of the respective sources, shown with a dashed black line.}
   \end{figure}

   Since the incidence angle $\theta$ is not known, fitting the fringes on the detector yields an effective optical thickness $\hat{D}=D\cos{(\theta)}$. Fig.~\ref{fig:fringe_fitting} shows the result of the fringe fitting for the point source and the extended source fringes. The parameters of interest are $\hat{D}$, as this will allow to determine a phase-shift between the two profiles, as well as the fitted fringe peak positions in wavenumber. We note that for the point source, fringes that cross a pixel column discontinuity are not fitted.
       
   In the second step, each point source fringe is linked to a single extended source fringe, which is in accordance with the picture shown in Fig.~\ref{fig:fringe_img_CVcampaign_zoom}. The phase-shift between a point source fringe $f_{ps}$ and its corresponding extended source fringe, $f_{ext}$, can be given by Eq. \ref{eq:phaseshift}:
   
       \begin{equation}\label{eq:phaseshift}
            \Delta\phi(f_{ext},f_{ps}) = 2\cdot \pi \cdot \hat{D}(f_{ext}) \cdot \left[\sigma(f_{ext} )-\sigma(f_{ps})\right]
       .\end{equation}
   
   The fitted fringe values are used for $\hat{D}(f_{ext})$, $\sigma(f_{ext})$, and $\sigma(f_{ps})$. The phase-shift is added to the sine part of Eq. \ref{eq:FPfunc}, as $\sin(2\pi \hat{D} \sigma + \Delta\phi)$.
       
   In the third step, similarly to the analysis in Sect.~\ref{sec:fringe_amplitude}, the distance at $f_{ps}$ from the PSF peak is computed and used as a baseline to compare different point source observations across the MRS FOV.
   
\begin{figure}[h]
   \centering
   \includegraphics[height=8.cm]{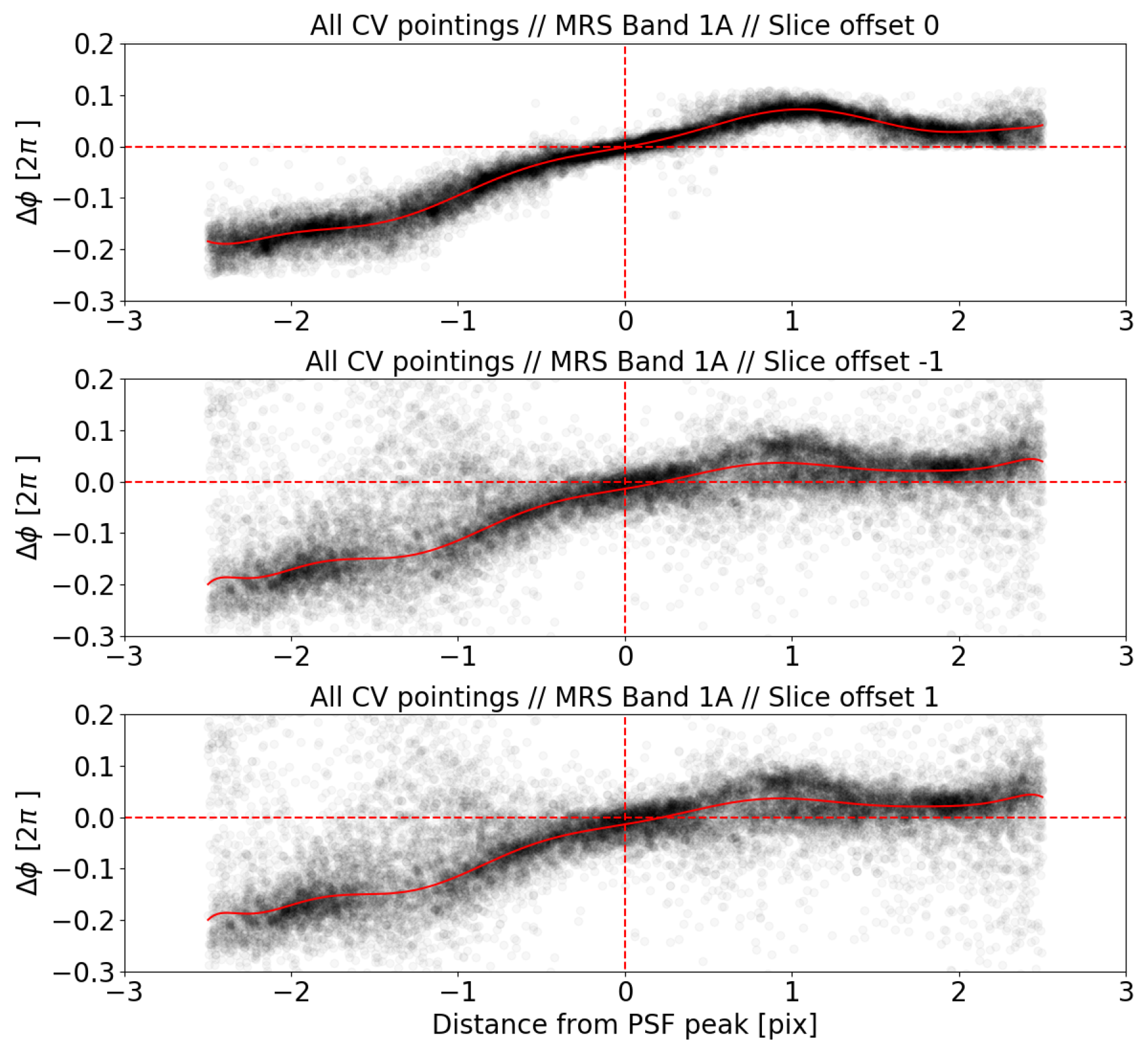}
   \caption[example]
   { \label{fig:phaseshift}
   Point source fringe phase as a function of distance from the PSF peak. A 12th order spline is fitted to the data (large uncertainties in slice offset -1 and slice offset +1).}
   \end{figure}

   The final result is shown in Fig.~\ref{fig:phaseshift}. Once again, we find that the systematic trend is smoothly varying and, similarly to what was found in Fig.~\ref{fig:psf_fringe_amplitude} and Fig.~\ref{fig:all_points_three_slices} for the fringe depth, the trend is not symmetric around the zero point. The low signal-to-noise ratio (S/N) in the neighboring slices (slice \textit{(s-1)} and \textit{(s+1)}) impedes the fringe fitting and the derivation of a clear phase-shift trend for those parts of the PSF.

    \begin{figure*}[h]
    \centering
    \includegraphics[height=9cm]{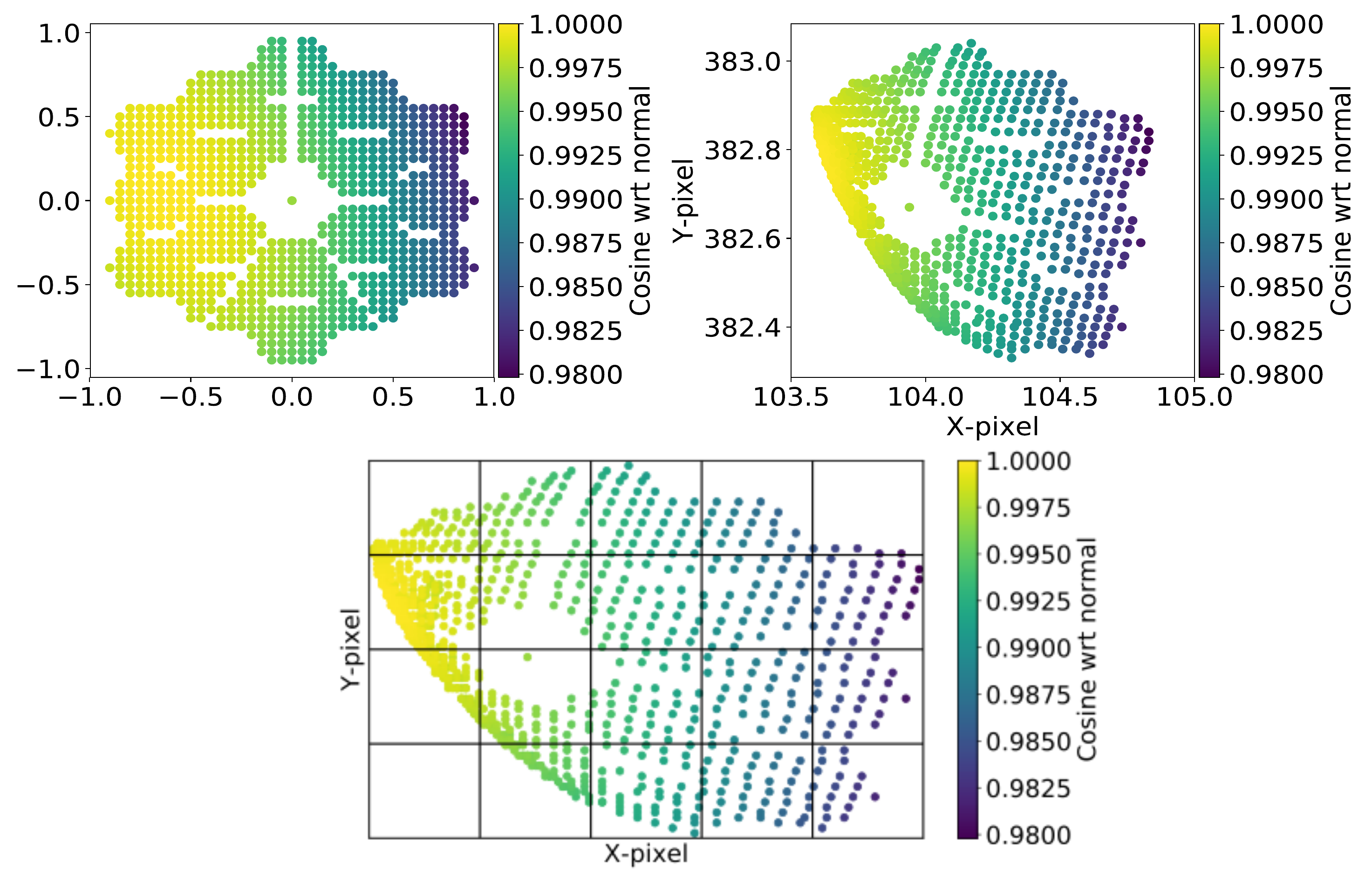}
    \caption[example]
    { \label{fig:pupil_illumination}
    Top: MIRI MRS angle of incidence cosine spot diagram computed using the Zemax model of the MIRI instrument. Bottom: Artificial 5-by-4 pixel grid overplotted on the detector spot diagram shown in the right plot of Fig.~\ref{fig:pupil_illumination}. This offers a simplistic view of the incidence angle distribution across the real MRS PSF.}
    \end{figure*}
    
    \begin{figure*}[h]
    \centering
    \includegraphics[height=8cm]{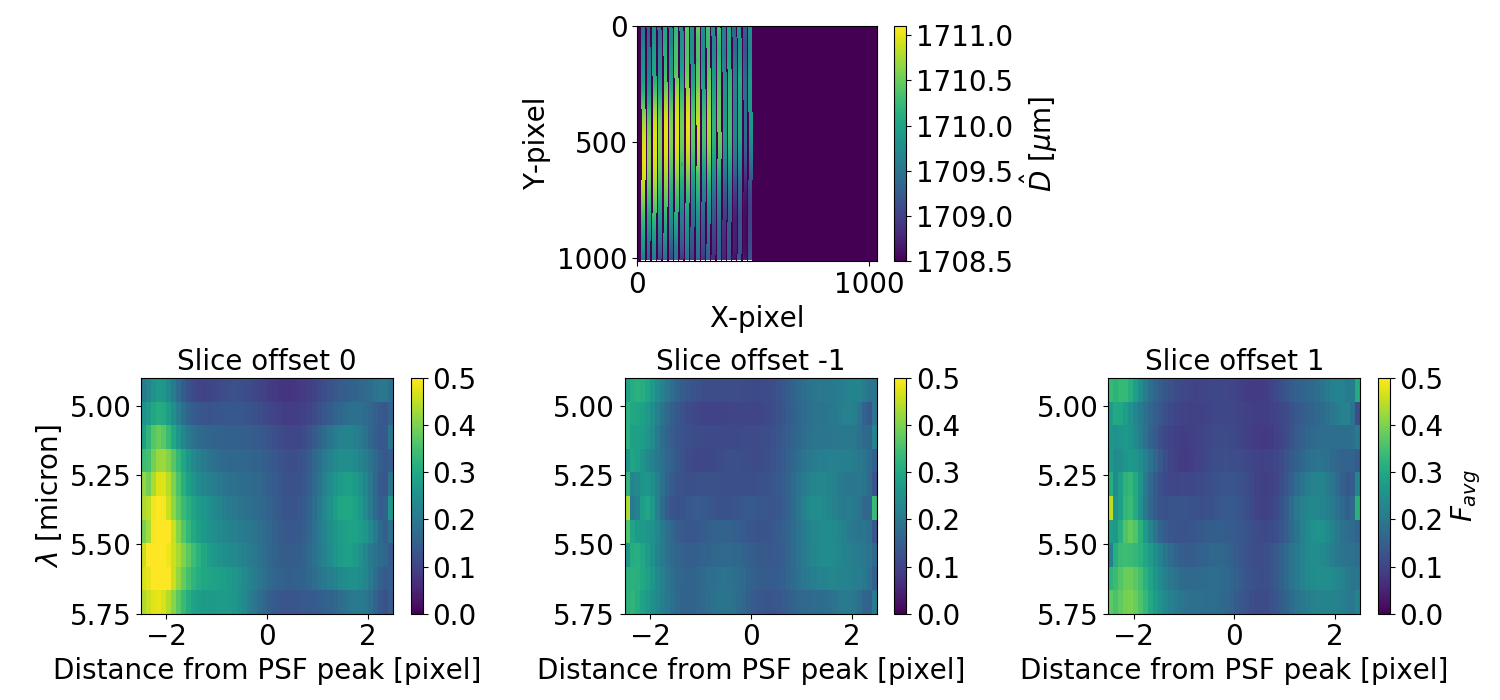}
    \caption[example]
    { \label{fig:FPtransm_model_parameters}
    Top row: Effective optical thickness $\hat{D}$ determined across MRS detector in band 1A. Bottom row: Average finesse $F_{avg}$ computed from the fringe depth trends shown in Fig.~\ref{fig:all_points_three_slices}.}
    \end{figure*}
   
   The three trends shown in Fig.~\ref{fig:phaseshift} display the same monotonic behavior as the trends shown in Fig.~\ref{fig:all_points_three_slices}. The fringe depth and the fringe phase are, in fact, connected. To explain this, we use Fig.~\ref{fig:pupil_illumination}, where we show the MRS angle of incidence cosine spot diagram of the MIRI pupil (top-left plot) on an MRS detector pixel (top-right plot). The cosine of the incidence angle with respect to the detector normal vector is derived from the Zemax optical model of the instrument. We note that Zemax modeling is purely geometric, i.e. it does not include diffraction. In the bottom plot of Fig.~\ref{fig:pupil_illumination}, we give a simplified view of the effect of diffraction on the MRS PSF, that is, the spot diagram gets bigger, covering, in this case,  a 4-by-5 detector pixel grid.
   
   Focusing on the bottom plot of Fig.~\ref{fig:pupil_illumination}, each pixel sees light at different incidence angles. This is illustrated by the color map. We note two things. Firstly, the phase of the fringes produced by the in-falling light changes with the cosine of the incidence angle. As a result, the final fringe phase recorded by a pixel will be proportional to the mean over all the fringe patterns produced by the incidence angles with which the light falls on the pixel. Secondly, integrating over different phases results in a blurring (reduction) of the fringe amplitude, thus a smaller fringe depth is measured.
  
  This new discovery allows us to reconcile the derived trends in fringe depth and fringe phase. Admittedly, a realistic simulation of the effects of diffraction on the Zemax MRS spot diagram would only marginally improve our understanding of point source fringes. Constraining the systematic effect of the illumination by the PSF and the effect of the varying incidence angle across the detector in the actual, as-built, MIRI MRS instrument can only be done empirically.
  
\subsection{New point source fringe correction method}
\label{sec:point_source_fringe_correction}
  We fold the systematic trends discovered in Sects. \ref{sec:fringe_amplitude} and \ref{sec:fringe_phase} into a new fringe correction for MIRI MRS spectra in band 1A. The correction operates by creating a fringe model based on Eq.~\ref{eq:FPfunc}. Since the exact (inhomogeneous) illumination pattern within a pixel is not known, we are limited to using average values for each parameter of Eq. \ref{eq:FPfunc}. There are three input parameters to the fringe model, (1) the effective optical thickness, $\hat{D}$, (2) an average finesse, $F_{avg}$, (3) the phase-shift, $\Delta\phi$.
  
  The optical thickness $\hat{D}$ can be mapped across the MIRI detector by fitting the extended source fringes in all the slices. We do so by scanning the fringes in the spectral direction. A fixed window is defined around every detector pixel and a fit is performed in that window before centering the window on the next pixel. The result is shown in the top row of Fig.~\ref{fig:FPtransm_model_parameters} for MRS band 1A.
  
  The average coefficient of finesse, $F_{avg}$, can be directly derived from the fringe depth shown in Fig.~\ref{fig:all_points_three_slices}. This is done by solving Eq. \ref{eq:FPfunc} for \textit{F} in the case that $\sin(2\pi \hat{D} \sigma) = 1$ and \textit{T$_e$} is equal to the fringe depth. We know that the fringe depth changes across the detector, as illustrated in Fig.~\ref{fig:all_points_three_slices_detdiffparts}. To map the values of $F_{avg}$, we constructed a 11-by-50 grid. The 11 vertical cells divide the wavelength range of the MRS band into 10 segments. For each of these segments, a 12th order spline is fitted to the values of $F_{avg}$ within that wavelength segment. The spline is then evaluated at 50 distances (distance from the PSF peak). The result is shown in the bottom row of Fig.~\ref{fig:FPtransm_model_parameters}.
  
  The phase-shift $\Delta\phi$ is determined from Fig.~\ref{fig:phaseshift}. Since no clear dependency was found between the phase-shift and the detector spectral direction in any of the three examined slices, a single 12th order spline is fitted to all the data at once, for each of the three slices examined.
  
  To correct a given point-source observation, the distance of each pixel to the PSF peak is computed in the three slices surrounding the PSF peak. Sequentially:
  \begin{itemize}
      \item the pixel position yields the effective optical thickness value at that pixel based on the optical thickness map shown in the top row of Fig.~\ref{fig:FPtransm_model_parameters}.
      \item Combining the distance information with the wavelength seen by the pixel, the average finesse is interpolated based on the grids shown in the bottom row of Fig.~\ref{fig:FPtransm_model_parameters}.
      \item The distance of the pixel to the PSF peak yields the phase-shift based on the splines shown in Fig.~\ref{fig:phaseshift}.
  \end{itemize} 
  
\begin{figure}[t]
  \centering
  \includegraphics[height=8.cm]{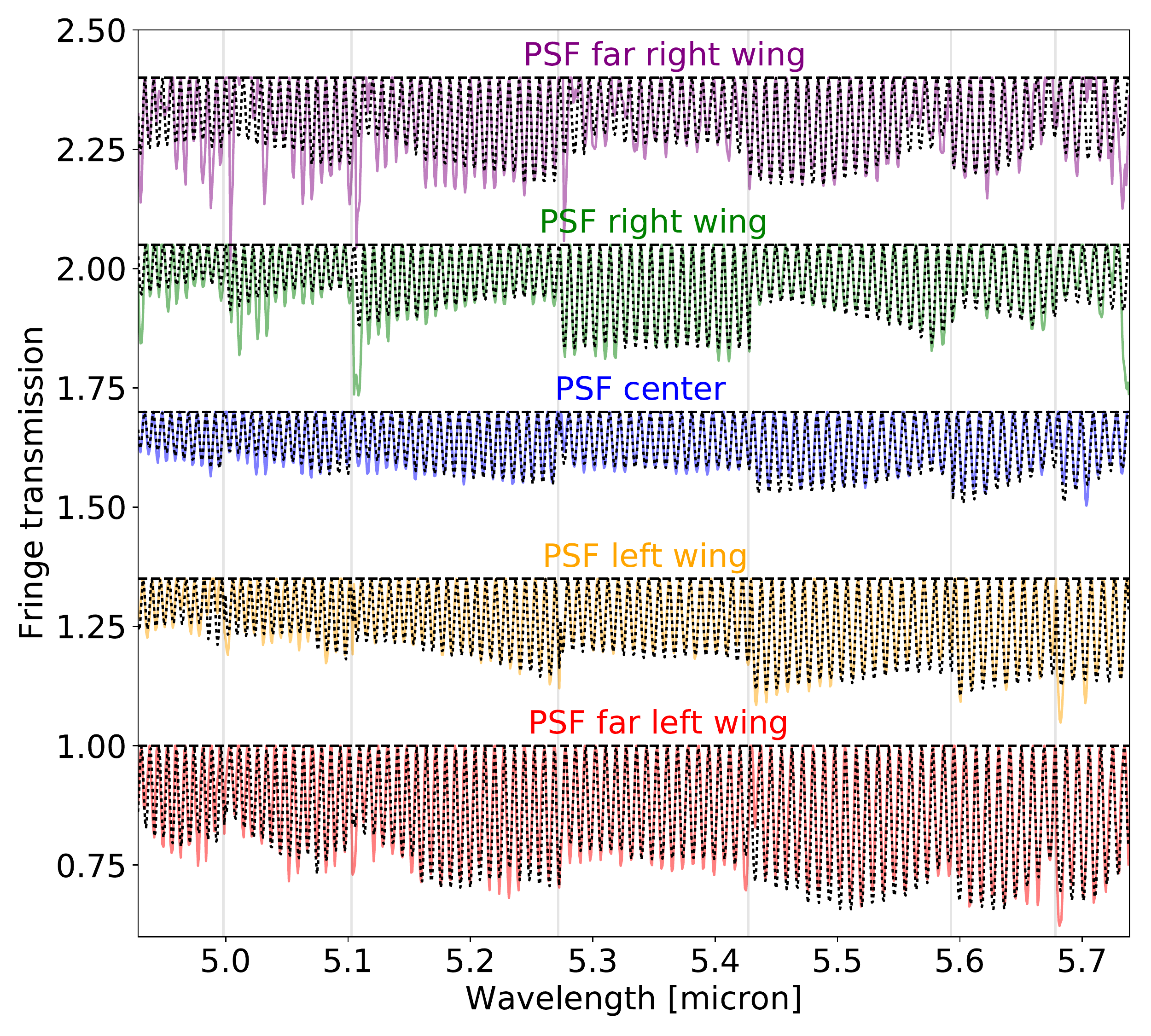}
  \caption[example]
  { \label{fig:FPtransm_vs_point_source}
  Point source fringe model (black dotted line) overplotted on fringe transmission determined empirically from a CV point source observation. Different isoalpha lines are offset vertically.}
  \end{figure}

  The above procedure yields a new point source fringe model at any location in the MRS FOV. Previously in Fig.~\ref{fig:ext_vs_point_source}, we compared a CV point source fringe transmission to that of an extended source observation. In Fig.~\ref{fig:FPtransm_vs_point_source}, we show the qualitative match between the point source fringe transmission and our new fringe model. The model matches the intricacies and jumps in the fringe depth and phase. 
  
\subsection{Spectral improvements with new fringe correction}
\label{sec:point_source_fringe_correction_improvement}
\subsubsection{Spectral continuum}
\begin{figure}[t]
  \centering
  \includegraphics[height=3.cm]{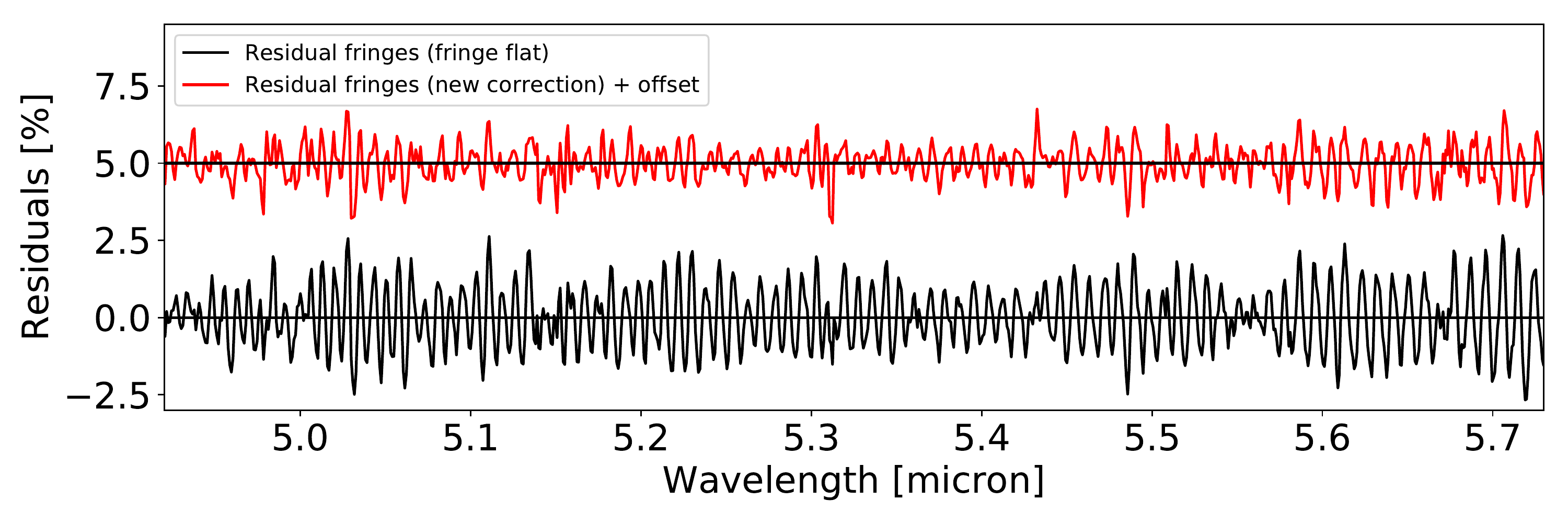}
  \caption[example]
  { \label{fig:residual_fringes_newcorrection}
  Residual fringes on point source integrated spectrum. The black data show the fringe residuals after dividing by the extended source fringe flat ($1\sigma\sim 1\%$). The red data, offset vertically by 5\%, show the fringe residuals after applying the new fringe correction ($1\sigma\sim 0.5\%$).}
  \end{figure}

  The effect of the new fringe model on the continuum of the point source integrated spectrum is shown in Fig.~\ref{fig:residual_fringes_newcorrection}. We see that the residual fringes have much less structure to them, as opposed to the application of the fringe flat based on an extended source observation. In the latter case, the residual fringes clearly show a remnant fringe modulation. The reported root-mean-square (RMS) errors on the two spectra are: $1\%$ in the case of the fringe flat correction (black data) and $0.5\%$ percent in the case of the new fringe correction (red data). Interestingly the phase-shift trends shown in Fig.~\ref{fig:phaseshift} display a counter-phase or anti-symmetry between the right and left side of the PSF. This implies that the fringes in the integrated spectrum are de facto smaller compared to looking at different parts of the PSF (by virtue of averaging sine waves in counter-phase). The new fringe correction accounts for this, minimizing correlated noise from summing different fringe profiles from different parts of the PSF.
  
\subsubsection{Line sensitivity}
\begin{figure}[t]
  \centering
  \includegraphics[height=9.cm]{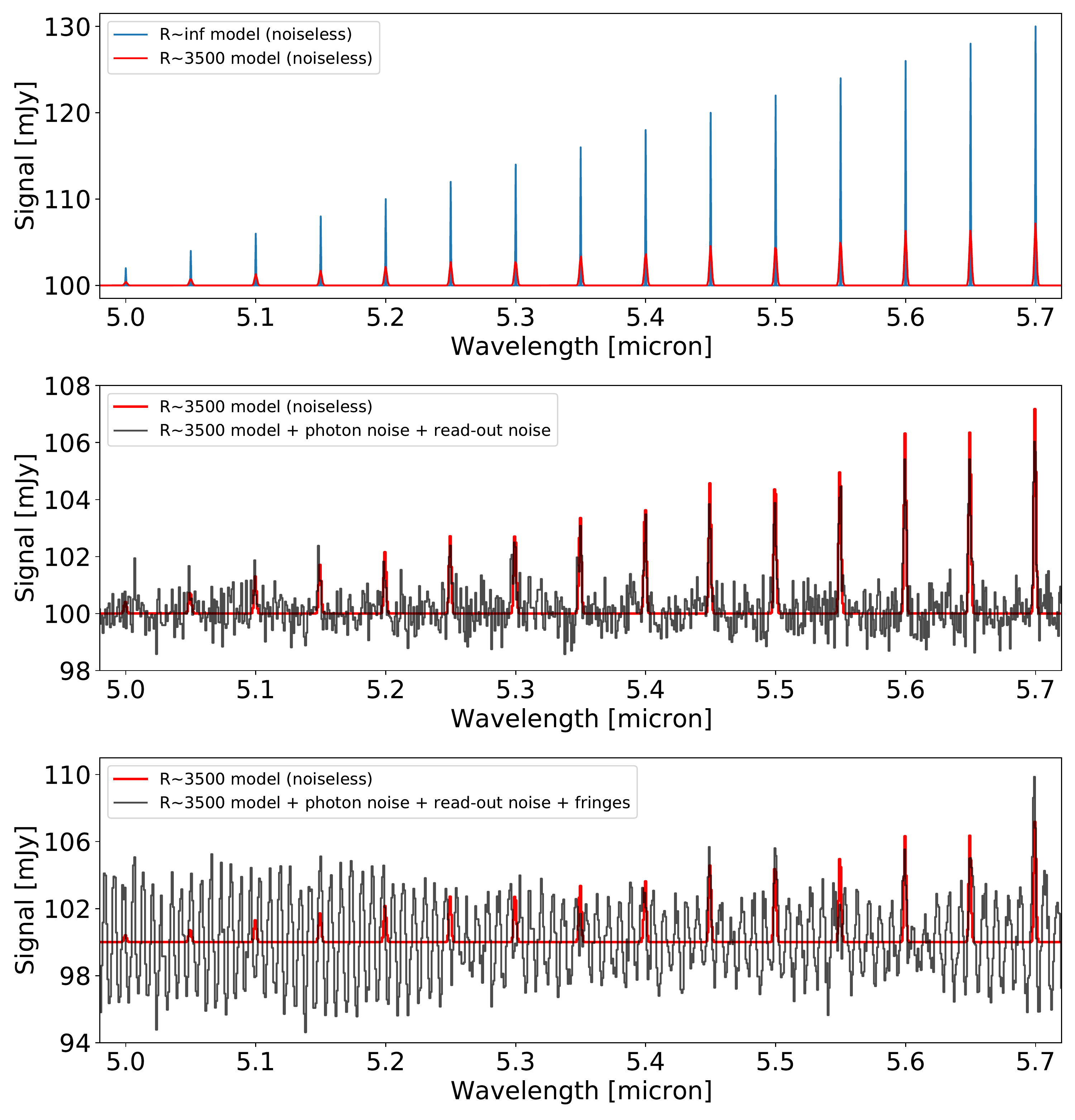}
  \caption[example]
  { \label{fig:line_sensitivity_test}
  Top plot: Initial model spectrum at two spectral resolutions. Middle plot: Degraded model spectrum with and without including photon noise and read-out noise. Bottom plot: Step plot of model spectrum at MIRI MRS resolution with photon noise, read-out noise, and fringes. The applied fringes are sourced from a CV integrated point source spectrum.}
  \end{figure}

  We showcase the improvement in the spectral line sensitivity using a fictitious spectrum with a constant spectral continuum of 100 mJy and distinct (unblended) spectral lines of varying intensity. Figure~\ref{fig:line_sensitivity_test} shows the test setup. The top plot shows our fictitious spectrum with an infinite spectral resolution.  The same spectrum degraded to the spectral resolution of the MIRI MRS in band 1A is overplotted. In the middle plot of Fig.~\ref{fig:line_sensitivity_test}, we show the noiseless (degraded) spectrum, as well as a spectrum that includes photon noise and read-out noise. The latter was produced using MIRISim, a public software package that produces realistic MIRI simulated data \citep{mirisim}. This noisy spectrum illustrates the point source sensitivity of the MRS in band 1A, specifically a 0.5mJy source achieves an SNR=10 after a 10.000 second on-source integration \citep{miri_pasp_9}. In the bottom plot of Fig.~\ref{fig:line_sensitivity_test}, we show the impact of the fringes when no correction is applied. We note that since MIRISim cannot simulate fringes for sources of different spatial sizes, we extracted the fringes from a CV point source integrated spectrum and applied them to the noisy spectrum of the middle plot, resulting in the bottom plot.
  
\begin{figure}[t]
  \centering
  \includegraphics[height=9.cm]{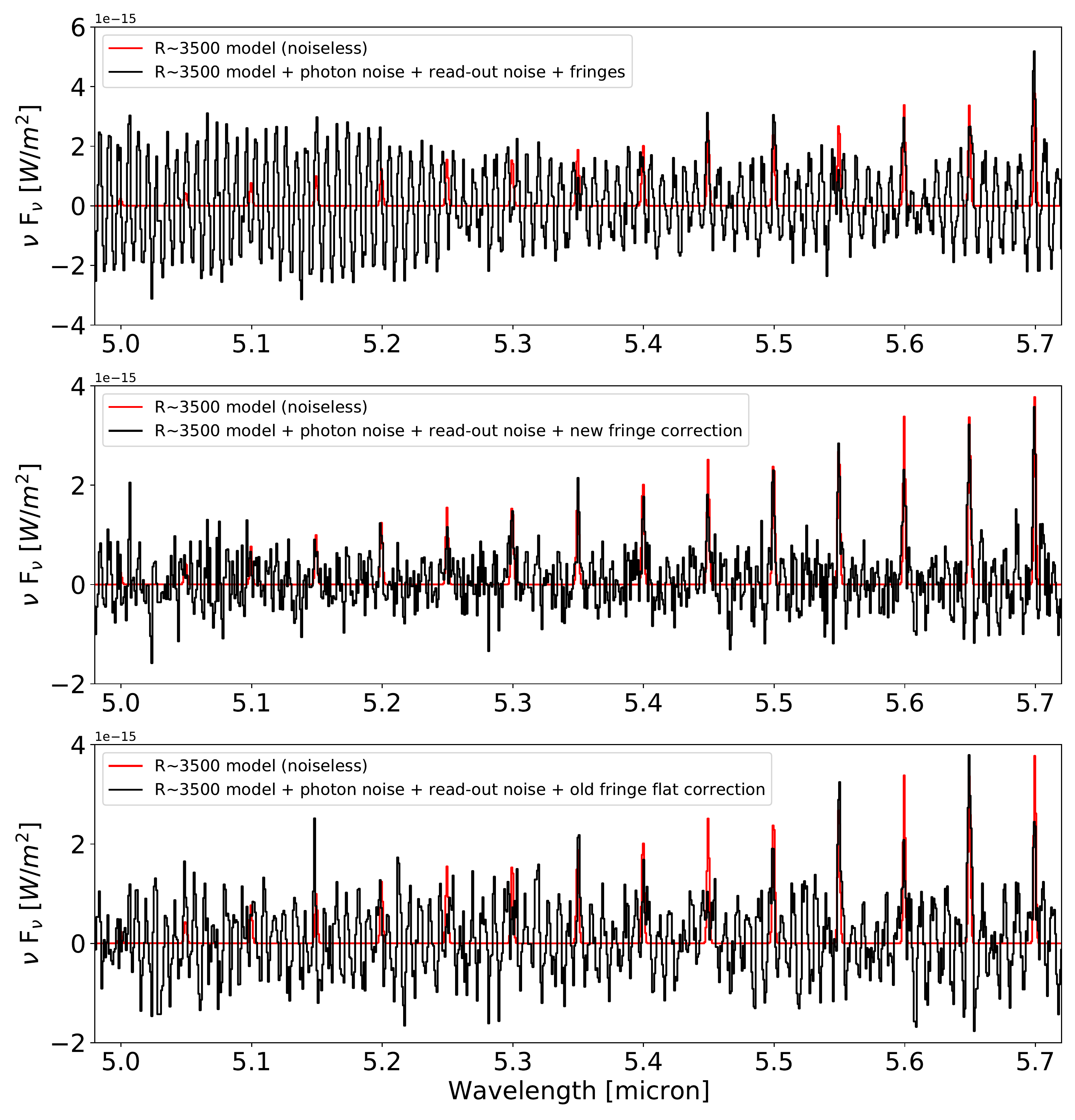}
  \caption[example]
  { \label{fig:line_sensitivity_test_fringes}
  Top plot: No fringe correction applied to model spectrum. Middle plot: New fringe correction applied to data. Weakest detectable line has a spectral irradiance of $1\cdot10^{-15} W/m^2$ (located at 5.2~$\mu m$). Bottom plot: Old extended source fringe flat applied to data. Weakest detectable line has a spectral irradiance of $2\cdot10^{-15} W/m^2$ (located at 5.35~$\mu m$).}
  \end{figure}

  In Fig.~\ref{fig:line_sensitivity_test_fringes}, we show the improvement in line sensitivity using the new point source fringe correction. First, we subtract the continuum of 100mJy from the bottom plot of Fig.~\ref{fig:line_sensitivity_test} and then we convert the signal units from Jansky to W/m$^2$; this is shown in the top plot of Fig.~\ref{fig:line_sensitivity_test_fringes}. The middle plot and the bottom plot show the result of using the new point source fringe correction and using the old correction with the extended source fringe flat, respectively. We report that with the new correction the weakest line we can detect has a spectral irradiance of $1\cdot10^{-15} W/m^2$ (located at 5.2~$\mu m$). Stronger lines are all detected. In the case of the old correction, the weakest detectable line has a spectral irradiance of $2\cdot10^{-15} W/m^2$ (positioned at 5.35~$\mu m$), indicating that the new correction detects lines that are half as strong.
  
  There are, nonetheless, things to keep in mind. Namely, since the fringes work in transmission, the stronger the continuum, the larger the impact of the fringes will be on the absolute signal. In addition, while we care about the standard deviation error for the continuum, with regard to the line sensitivity, it is the peak spectral variation that matters at any given wavelength. Lastly, the impact of the fringes is not uniform across the wavelength range (the modulation varies across the spectrum), hence, it is difficult to comment on a global improvement. The location where a line falls with respect to the fringe peaks or troughs plays a role in the detectability of a spectral line.
  
\subsubsection{Shape of weak spectral lines}
  In order to provide some insight on the improvement of the shape of weak spectral lines after fringe correction, we use a T Tauri SED model; T Tauri protoplanetary discs are rich in spectral lines in the mid-infrared. The model used is based on the numerical output of the ProDiMo code by \cite{t_tauri_diana}. We translated the publicly available ProDiMo IDL code to Python and constructed a model from the default T Tauri SED by multiplying the entire spectrum (star and disc) by a fudge factor of 10. This was necessitated by the fact that due to the small line fluxes, after degrading the original (not fudged) spectrum to the MRS spectral resolution and adding photon noise and read-out noise, most of the lines could not be detected.

\begin{figure*}[h]
  \centering
  \includegraphics[height=6.1cm]{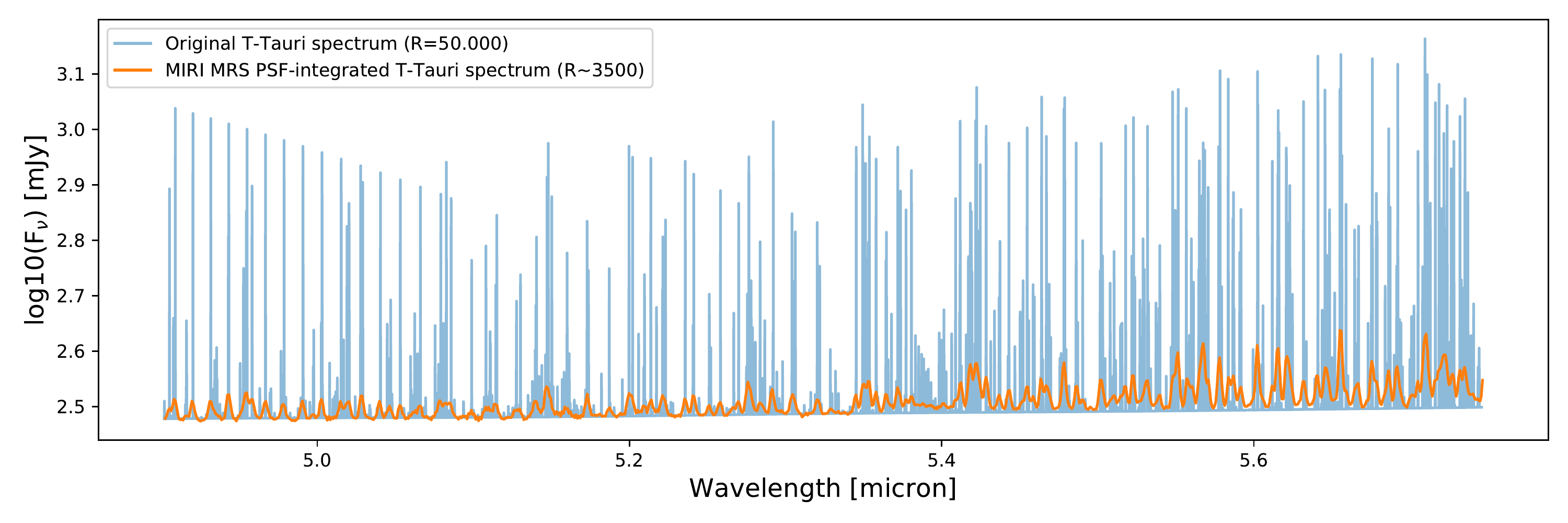}
  \caption[example]
  { \label{fig:t_tauri_miri}
  Comparison between a T Tauri spectrum constructed using the numerical output of the ProDiMo code and fudged by a factor of 10, and a PSF-integrated MIRI MRS spectrum of the same T Tauri spectrum. For this figure, no noise contributions were added to the MIRI simulation.}
  \end{figure*}

  We used MIRISim to simulate an unresolved source with the fudged T Tauri SED model as input. Fig.~\ref{fig:t_tauri_miri} shows a comparison between the input high-resolution T Tauri spectrum and the PSF-integrated MIRI MRS spectrum in band 1A. A single exposure is simulated with a single integration of 200 seconds. In Fig.~\ref{fig:t_tauri_miri} no noise contributions were added to the MIRI simulation, thus, the integration time is of no consequence. A prominent blending of lines can be distinguished in the MIRI spectrum, which illustrates what can be expected from observing a T Tauri star with the MRS.
  
  \begin{figure*}[h]
  \centering
  \includegraphics[height=12.1cm]{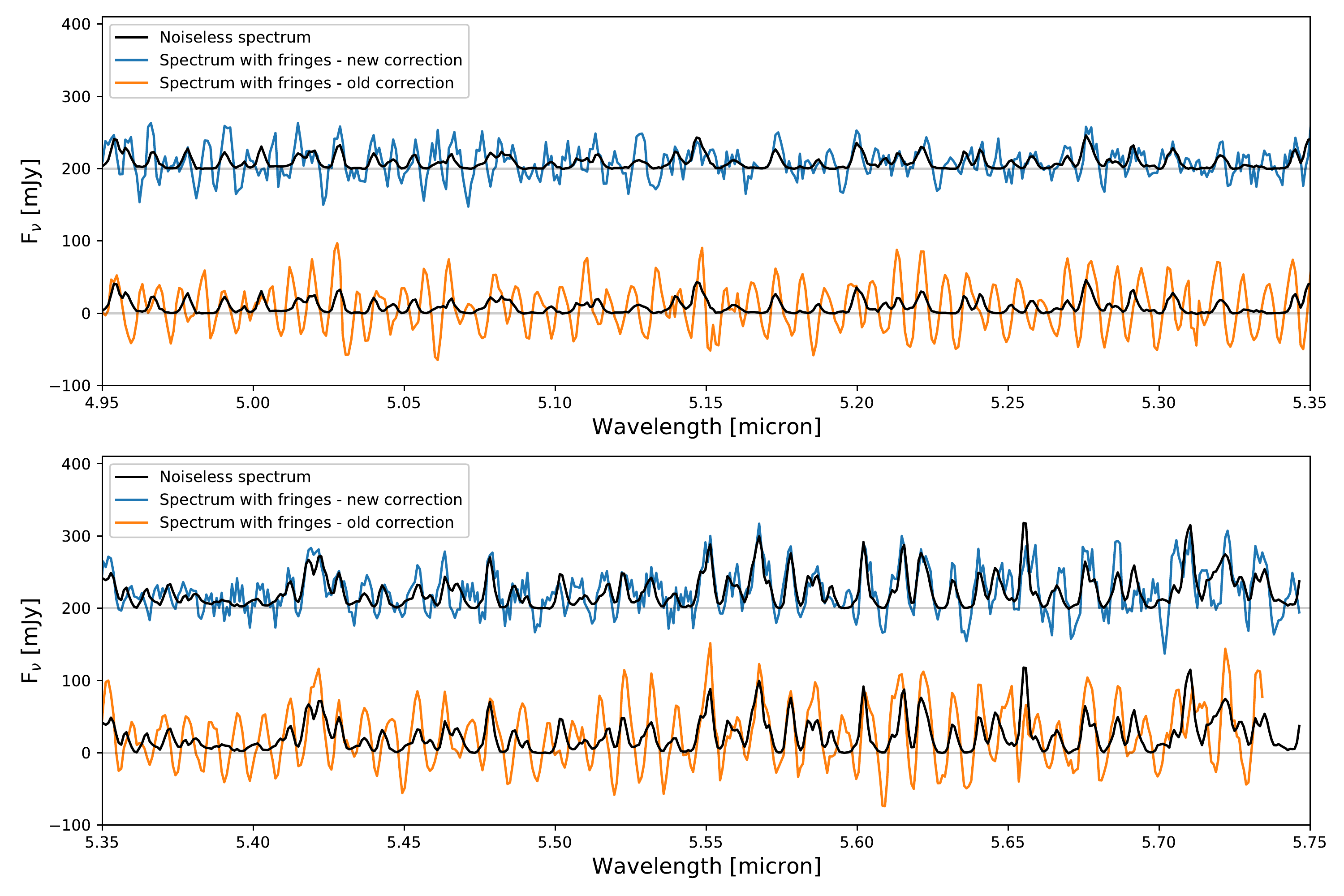}
  \caption[example]
  { \label{fig:t_tauri_spectrum_fringes}
  Comparison in line shapes of T Tauri spectrum using the new point source fringe model versus using the old extended source fringe flat. A vertical offset of 200mJy is applied between the two cases, the noiseless spectrum (black line) is the same in both cases. The two plots cover one half of the band 1A wavelength range. On average the new fringe correction does a better job at returning the original spectral line shapes.}
  \end{figure*}
  
  The improvement in the weak spectral line shape is illustrated qualitatively in Fig.~\ref{fig:t_tauri_spectrum_fringes}. Here, we added photon noise and read-out noise to the T Tauri spectrum, as well as introduced fringes and subtracted the known (noiseless) SED continuum. Figure~\ref{fig:t_tauri_spectrum_fringes} shows the impact of using the new point source fringe correction versus the extended source fringe flat. On average, the new fringe correction does a better job at retrieving the original spectral line shapes.
  
\subsection{From point source fringes to extended source fringes}
\label{sec:from_point_to_extended}
   An extended source can be seen as a collection of point sources. In this section, we show that an extended source behaves like the sum of point sources also with respect to the fringes. 
   
   In Fig.~\ref{fig:fringe_img_CVcampaign_zoom}, we overplotted the point source fringe transmission on top of the extended source fringe transmission. There is a clear overlap between the fringe peaks of the point source (bright regions) and the fringe peaks of the extended source. Since an extended source is equivalent to a superposition of many point sources, this means that the extended source fringes must be the result of averaging multiple point source fringes in the spatial (X-pixel) direction. 
   
\begin{figure}[h]
   \centering
   \includegraphics[height=3.cm]{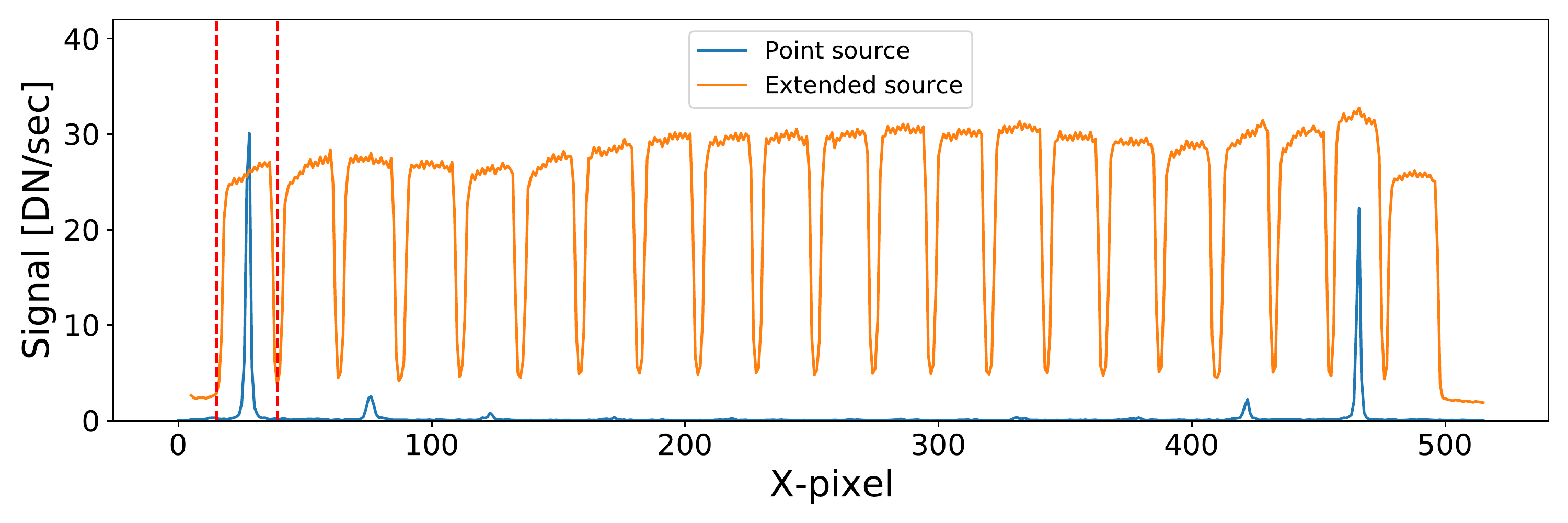}
   \caption[example]
   { \label{fig:use_psf_core}
   Signal of point source and extended source observed with the MRS in a single detector row.}
   \end{figure}

\begin{figure}[h]
   \centering
   \includegraphics[height=3.5cm]{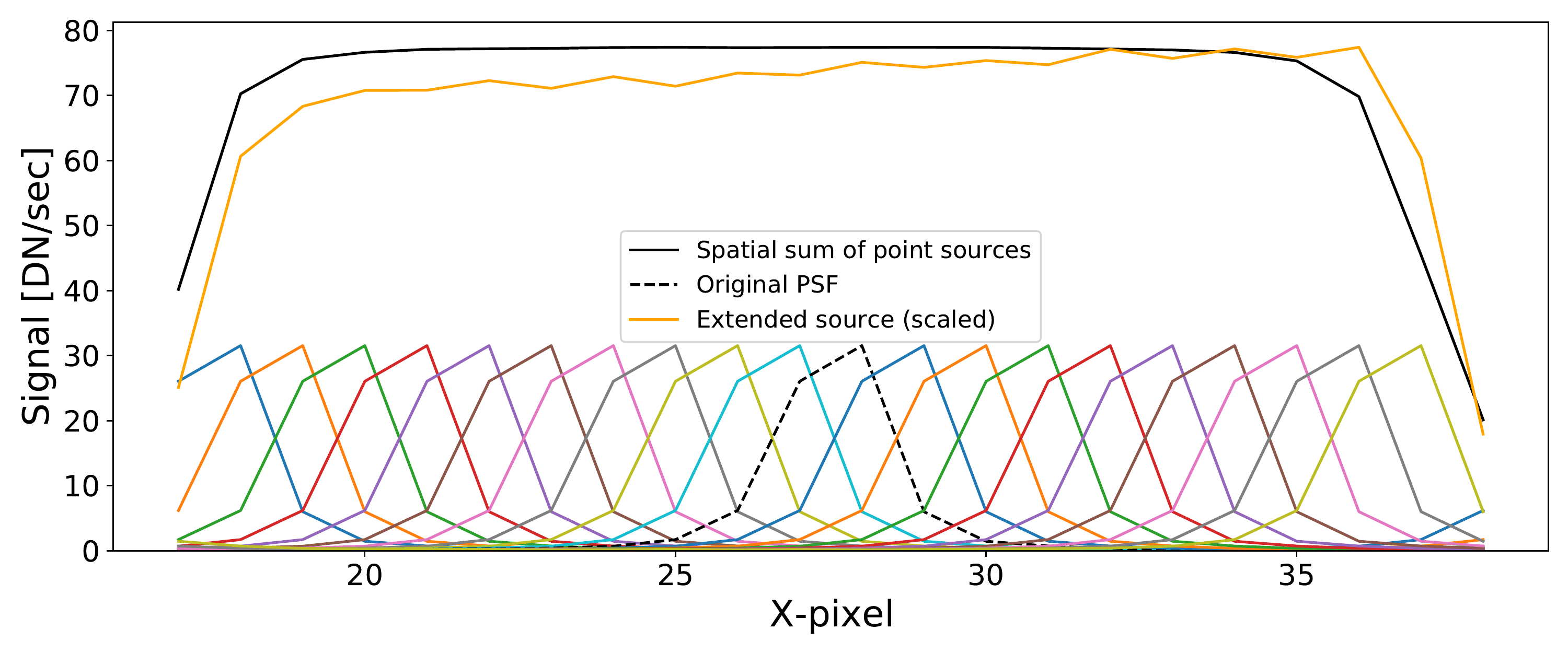}
   \caption[example]
   { \label{fig:roll_psf}
   A point source (dashed black line) is rolled by one pixel and duplicated at different X-pixel locations. The point source profiles are then summed to yield a flat-topped illumination. This is shown by the black continuous line. A real extended source observation with the MRS is overplotted for comparison.}
   \end{figure}

   We use the MIRI extended source data and a point source observation from the CV campaign. In Fig.~\ref{fig:use_psf_core}, the two sources are overplotted in one detector row. To simulate the presence of multiple point sources in a slice, we use the PSF (blue curve) between the vertical dashed red lines in Fig.~\ref{fig:use_psf_core}. In Fig.~\ref{fig:roll_psf}, we illustrate the PSF with the black dashed profile. We roll the profile by one pixel to the left and one pixel to the right until the full slice is filled. We then sum the duplicated profiles. This is equivalent to convolving the point source spectrum with a box function, meaning that it yields a flat-topped illumination. The (scaled) extended source signal is shown for comparison.
   
\begin{figure}[h]
   \centering
   \includegraphics[height=7.5cm]{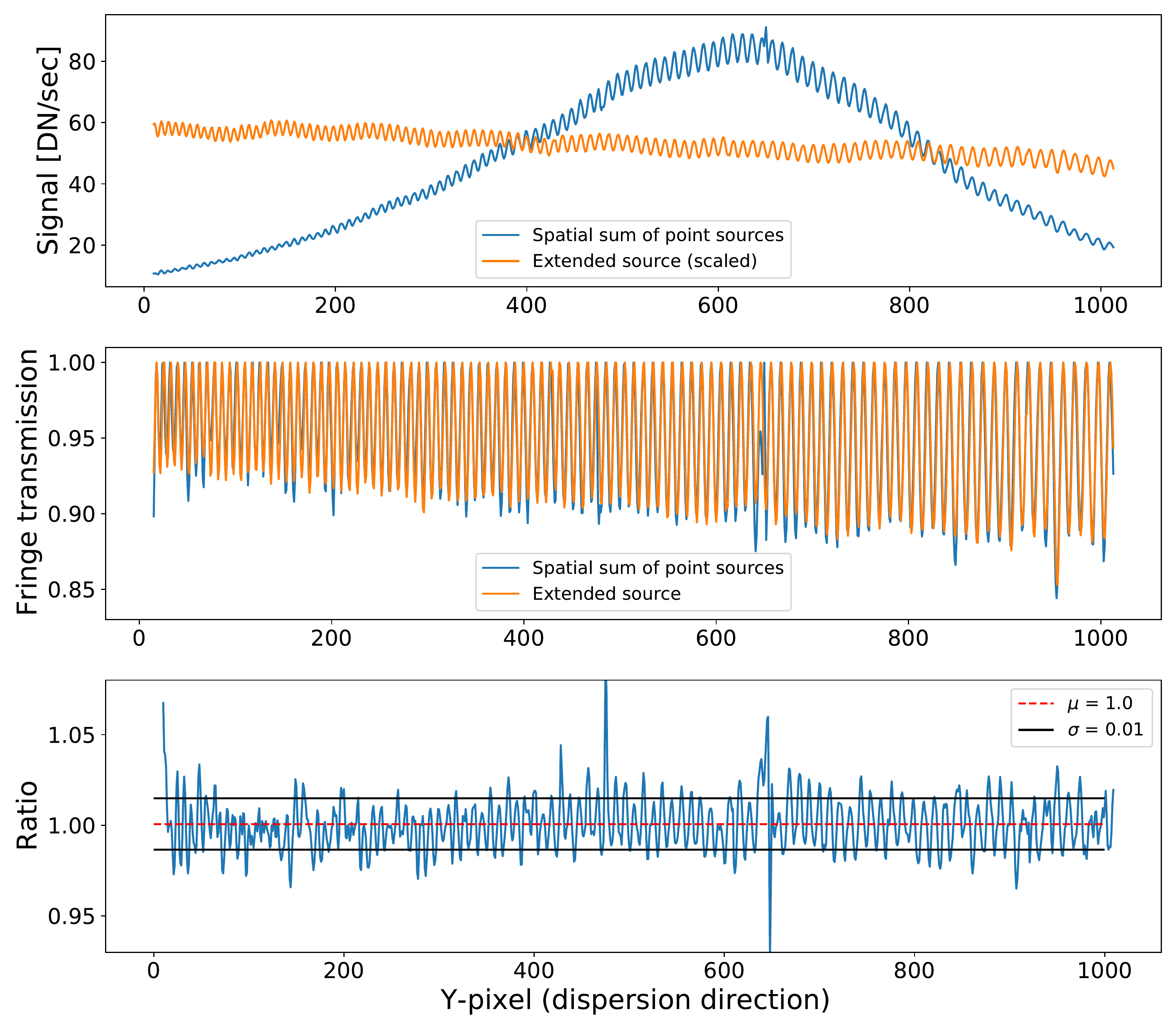}
   \caption[example]
   { \label{fig:point_to_extended_comparison}
   Top plot: Comparison between FM extended source spectrum and spectrum derived from summing multiple point sources in the MRS FOV. The two spectra have a different spectral baseline, caused by the different spectral energy distribution of the sources. Middle plot: Comparison of the fringe transmission derived from analyzing the spectra in the top plot. Bottom plot: Ratio of the fringe transmission profiles of the middle plot. The mean and standard deviation of the ratio are delineated with the red dashed line and the black continuous lines, respectively.}
   \end{figure}

   The procedure shown in Fig.~\ref{fig:roll_psf} is performed for all MRS detector rows. This means that an extended source is effectively simulated in both the spatial and spectral direction. We show the comparison between the real extended source and the simulated extended source in the top plot of Fig.~\ref{fig:point_to_extended_comparison}. The two spectra are evaluated at the centroid of the CV pointing. In order to better visualize how the fringes in the two spectra compare to each other, we normalize the spectra to the sampled fringe peaks of the respective spectra. This results are displayed in the middle plot of Fig.~\ref{fig:point_to_extended_comparison}. To quantify the similarity between the two fringe transmission profiles, the ratio of the two is computed. This is illustrated in the lower plot of Fig.~\ref{fig:point_to_extended_comparison}. We compute a 1-sigma standard deviation of one percent.
   
   To get the result of Fig.~\ref{fig:point_to_extended_comparison} a number of simplifications are made: 1) The roll of the PSF was done in a horizontal direction, instead of following the curvature of the isolambda lines; 2) a pixel-sampled point source observation was used; 3) The sampled PSF was rolled by discrete single pixel steps; 4) the duplicated PSFs were all given the same weight.
   
\subsection{Fringe correction for complex scenes}
\label{sec:semi_extended_source_correction}
\subsubsection{Semi-extended sources}
  \begin{figure*}[h]
  \centering
  \includegraphics[height=9cm]{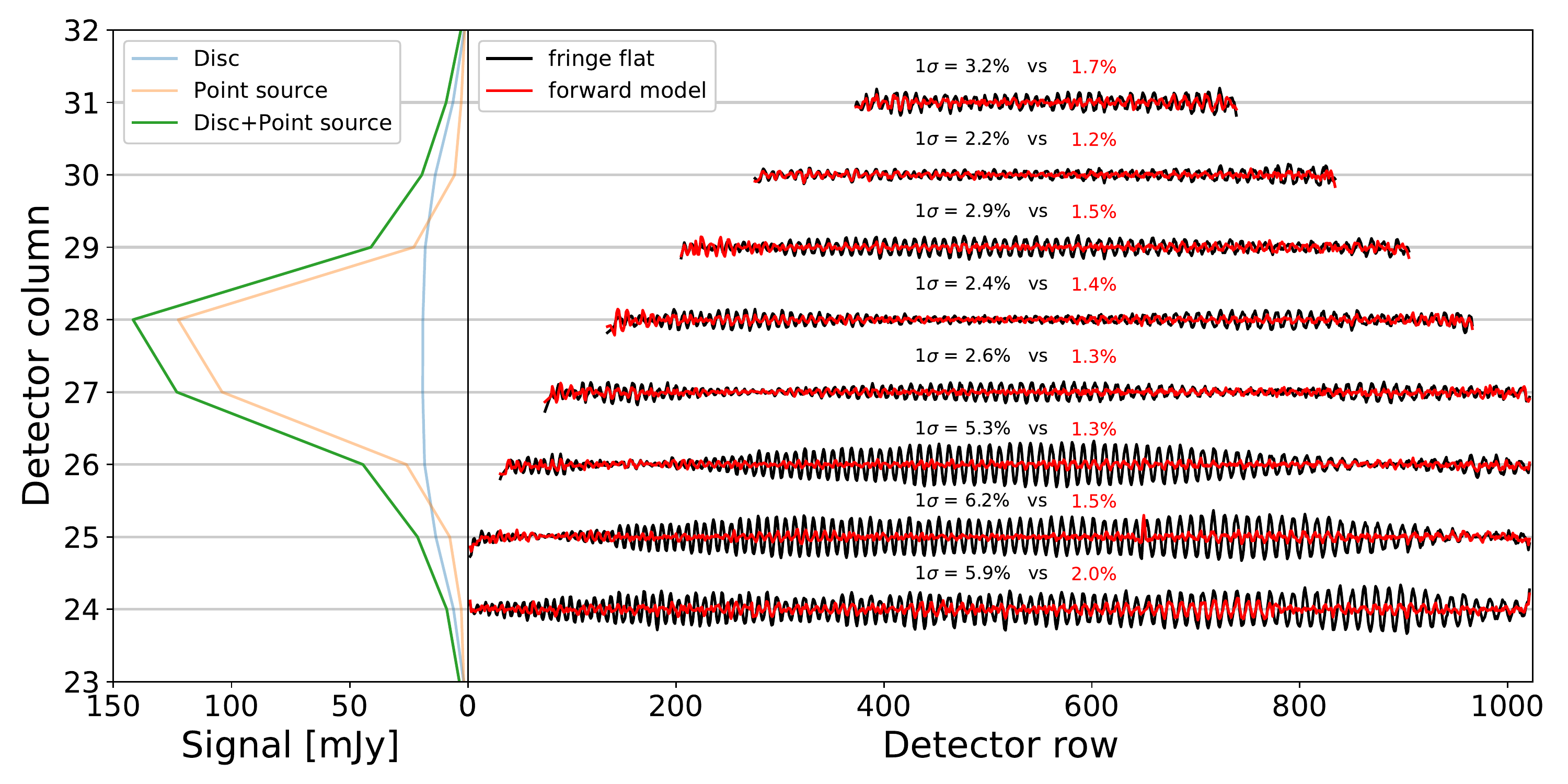}
  \caption[example]
  { \label{fig:point_source_and_disc_residuals}
  Left: Simulated scene with point source and semi-extended source. Right: Comparison in fringe residuals after using a single fringe flat based on an extended source observation (black data) versus using a fringe model that matches the spatial distribution of the sources (red data).}
  \end{figure*}
  
  \begin{figure*}[h]
  \centering
  \includegraphics[height=9cm]{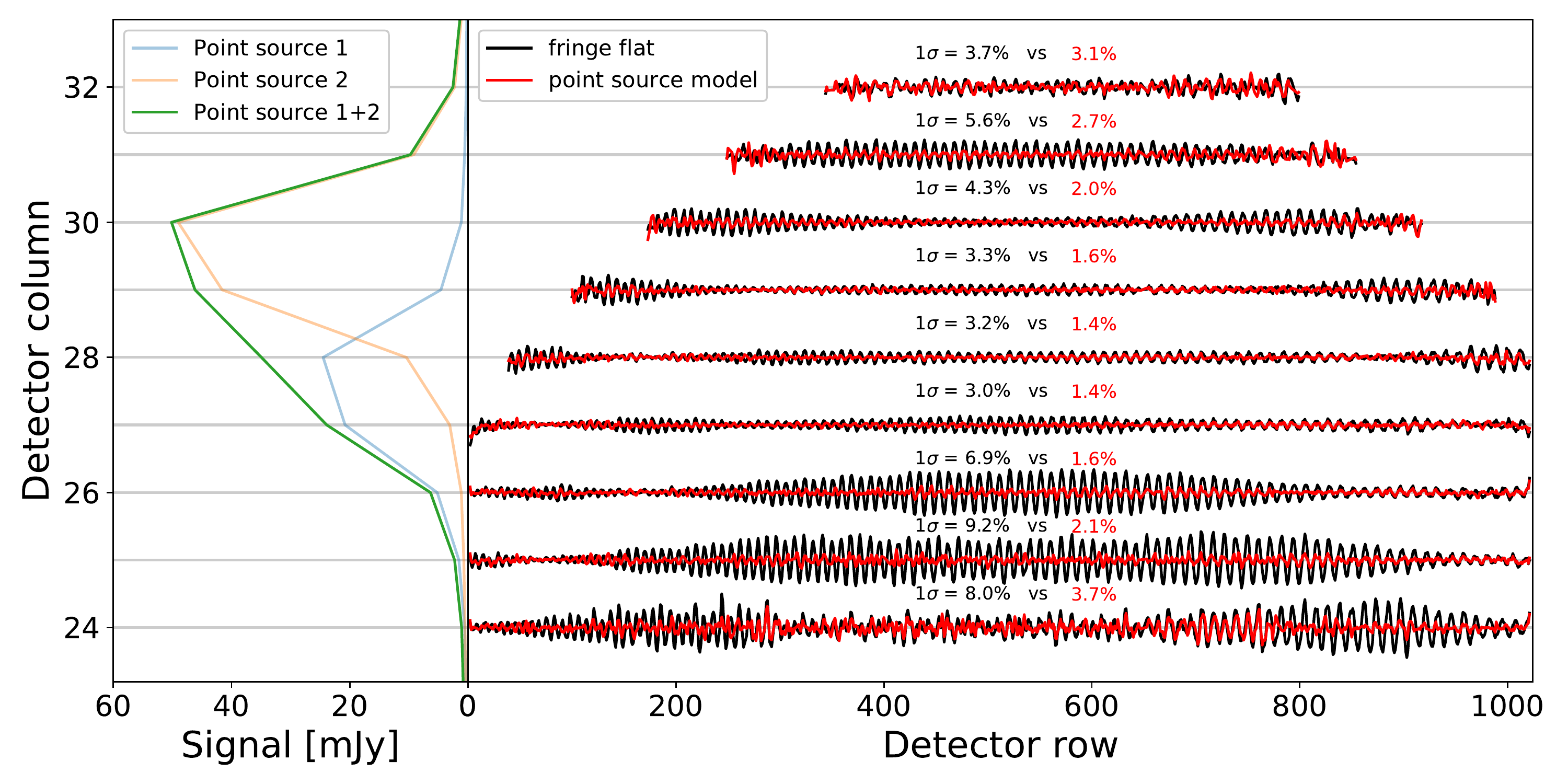}
  \caption[example]
  { \label{fig:crowded_field_residuals}
  Left: Scene with two nearby CV point sources. Right: Comparison in fringe residuals after using a single fringe flat based on an extended source observation (black data) versus using a point source fringe model for each point source in the MRS FOV (red data).}
  \end{figure*}

   Despite the very simple approach and its limitations, we achieve an impressively high level of accuracy in reproducing the extended source fringes. This result strongly suggests that the point source fringe model can be used to correct for the fringes in any semi-extended source observation by forward modeling. For this, we need to model the spatial distribution of the flux received from the source. This could be done based on MIRI imaging observations, or a spectrally collapsed pre-processing of the MRS observation itself. The point source fringe flat model can then be convolved by the scene and the result is divided out of the data. 
   
   In Fig.~\ref{fig:point_source_and_disc_residuals}, we show an example of a simulated scene containing a point source and a semi-extended source. The semi-extended source is created by convolving the point source with a six pixel wide box kernel. The scene geometry and signal, as seen in one row and one slice on the detector, is shown in the left panel of the figure, where we sum the response-corrected signal of the two sources.
   
   We identify two fringe calibration cases, an "old" calibration, which involves dividing by the extended source fringe flat, and a "new" calibration based on a forward model of the scene, which can be informed using the data themselves. The two cases are described mathematically by Eq.~\ref{eq:pipeline_correction} and Eq.~\ref{eq:forward_model}. In Eq.~\ref{eq:pipeline_correction}, "F$_{ext}$" stands for the extended source fringe flat. The mathematical operation in the denominator of Equ.~\ref{eq:forward_model} is a convolution between the normalized scene flux, and the point spread function of the MRS that includes the point source fringe model.

  \begin{equation}\label{eq:pipeline_correction}
     S_{cal,old} = S_{tot} / F_{ext}
  ,\end{equation}
  
  \begin{equation}\label{eq:forward_model}
     S_{cal,new} = \frac{S_{tot}}{ \left(\frac{S_{tot}}{\sum S_{tot}}\right) * PSF_{w/fringes}}
  .\end{equation}
  
  In the case of having a scene with a point source and a disc, $S_{tot} = S_{ps} + S_{disc}$, where $S_{tot}$ is the flux signal measured on the detector. In the right panel of Fig.~\ref{fig:point_source_and_disc_residuals}, we show the fringe residuals in the detector's spectral (column) direction based on the two different calibration schemes. The curvature of the orders (isoalpha lines) on the detector means that the spectrum extends over less rows in some columns than in others. We omit pixels with signal smaller than 1mJy, as those are dominated by noise.
  
  We know that point source fringes are most similar to extended source fringes near the peak of the PSF. This is illustrated in the residual fringes. The forward model accounts for the systematic nature of the fringe variation across the PSF, hence, it does a more consistent job in correcting the fringes at all points in the field.
  
\subsubsection{Crowded fields}
   The point source fringe model can also be used to correct the fringes measured in a scene with a crowded field. The only requirement to make a forward model in this case is to know the centroid of each source. Several PSFs comprising the model fringes can then be summed together before normalizing the flux distribution. Putting this in the context of Eq.~\ref{eq:forward_model}, in this case $S_{tot} = S_{ps_1} + S_{ps_2} + ... + S_{ps_N}$.
  
  An example of a crowded field is shown in Fig.~\ref{fig:crowded_field_residuals}. In the left panel, we sum the contribution of two point sources based on the CV pointings. The data are response-corrected. In the right panel, we show the fringe residuals using the two calibration approaches described by Eq.~\ref{eq:pipeline_correction} and Eq.~\ref{eq:forward_model}. In the latter case, we record standard deviation values about half of the former case. Furthermore, the fluctuations in the residuals across the field are minimized.
  
  The forward modeling methods presented in this section can be translated to other infrared instruments. The method requires to have a good knowledge of the instrument point spread function and to know how the fringe parameters vary as a function of the given part of the PSF that is sampled. This combination can then be used as a numerical kernel to model the fringes in any complex scene.
   
\section{Conclusions}
\label{sec:conclusions}
   In this paper, we show that fringes in point source observations display important variability in amplitude and phase when sampling different parts of the MRS PSF on the detector. We carried out a closer study of point source fringes to explain this effect.
   
   By collapsing multiple ground-based observations of point sources across the MRS FOV, we show that the fringe depth varies smoothly as a function of the fraction of the PSF falling within a pixel. This means that the analytic description of MRS fringes for point sources is the same over the FOV at any given wavelength and depends on the part of the PSF that falls within a pixel.
   
   Furthermore, we discovered a reproducible linear trend in the point source fringe phase. The trends in the fringe phase and fringe depth are linked by virtue of the diffraction-broadened MRS pupil illumination being asymmetric and yielding a non-uniform spread of incidence angles on each detector pixel. The fringe phase is found to be in counter-phase in the two sides of the PSF. This was determined by using the fringes of an extended, spatially homogeneous source as reference.
   
   In combining all the new quantitative information on the fringe depth and the fringe phase, we present a novel method to derive a fringe flat applicable to a point source observed with MIRI MRS at any location of the MRS FOV. We verify the method on a random CV point source. The new correction yields a 50$\%$ smaller 1$\sigma$ standard deviation in the continuum. An improvement of 50$\%$ in line sensitivity was determined based on a benchmark test with a continuum of 100mJy. We also show that the new correction allows for an easier retrieval of the spectral lines of a T Tauri spectrum as observed with MIRI.
   
   Finally, we prove that the fringe profile of an extended source can be derived from a point source observation. Despite the simple approach applied in this work, we achieve an accuracy of 1\%\ RMS in reproducing the extended source fringes. The proposed algorithm could be used to model the fringes in sources with spatial structure, as demonstrated in this paper. For crowded fields, a fringe flat can be produced by knowing the centroid position of each source in the field. This forward modeling method, which uses a combination of the point spread function and the fringes in every part of the point spread function as a kernel, can be translated to other infrared instruments.
   
   Understanding the different components of MIRI calibration that impact the trends derived in this paper will allow us to better constrain the analytical expressions derived for the fringe depth and fringe phase. These could then be used to correct for the fringes in any point source observation and in a  crowded field, as well as correct the fringes in any extended source with spatial structure by forward modeling. While we caution that the data at our disposal only allow us to test this method in the 1A MRS band, we do expect it to be applicable to point sources as well as to extended sources in bands 1B to 4C. Several point sources with a bright spectral continuum from 4.8~$\mu m$ to 28.8~$\mu m$ will be observed during the MIRI instrument commissioning phase. These will then be used to derive a point source fringe correction in all of the MRS spectral bands.
   
\begin{acknowledgements}
   Ioannis Argyriou, Pierre Royer, and Bart Vandenbussche thank the European Space Agency (ESA) and the Belgian Federal Science Policy Office (BELSPO) for their support in the framework of the PRODEX Programme. \\
   Patrick Kavanagh thanks the European Space Agency (ESA) and Enterprise Ireland for their support in the framework of the PRODEX Programme. \\
   Alvaro Labiano acknowledges the support from Comunidad de Madrid through the Atracción de Talento grant 2017-T1/TIC-5213.\\
   Ioannis Argyriou also thanks Clio Gielen (KU Leuven, Belgium) for many useful and stimulating discussions.\\
   The work presented is the effort of the entire MIRI team and the enthusiasm within the MIRI partnership is a significant factor in its success. MIRI draws on the scientific and technical expertise of the following organisations: Ames Research Center, USA; Airbus Defence and Space, UK; CEA-Irfu, Saclay, France; Centre Spatial de Liége, Belgium; Consejo Superior de Investigaciones Científicas, Spain; Carl Zeiss Optronics, Germany; Chalmers University of Technology, Sweden; Danish Space Research Institute, Denmark; Dublin Institute for Advanced Studies, Ireland; European Space Agency, Netherlands; ETCA, Belgium; ETH Zurich, Switzerland; Goddard Space Flight Center, USA; Institute d'Astrophysique Spatiale, France; Instituto Nacional de Técnica Aeroespacial, Spain; Institute for Astronomy, Edinburgh, UK; Jet Propulsion Laboratory, USA; Laboratoire d'Astrophysique de Marseille (LAM), France; Leiden University, Netherlands; Lockheed Advanced Technology Center (USA); NOVA Opt-IR group at Dwingeloo, Netherlands; Northrop Grumman, USA; Max-Planck Institut für Astronomie (MPIA), Heidelberg, Germany; Laboratoire d’Etudes Spatiales et d'Instrumentation en Astrophysique (LESIA), France; Paul Scherrer Institut, Switzerland; Raytheon Vision Systems, USA; RUAG Aerospace, Switzerland; Rutherford Appleton Laboratory (RAL Space), UK; Space Telescope Science Institute, USA; Toegepast- Natuurwetenschappelijk Onderzoek (TNO-TPD), Netherlands; UK Astronomy Technology Centre, UK; University College London, UK; University of Amsterdam, Netherlands; University of Arizona, USA; University of Bern, Switzerland; University of Cardiff, UK; University of Cologne, Germany; University of Ghent; University of Groningen, Netherlands; University of Leicester, UK; University of Leuven, Belgium; University of Stockholm, Sweden; Utah State University, USA. A portion of this work was carried out at the Jet Propulsion Laboratory, California Institute of Technology, under a contract with the National Aeronautics and Space Administration.

   We would like to thank the following National and International Funding Agencies for their support of the MIRI development: NASA; ESA; Belgian Science Policy Office; Centre Nationale D'Etudes Spatiales (CNES); Danish National Space Centre; Deutsches Zentrum fur Luft-und Raumfahrt (DLR); Enterprise Ireland; Ministerio De Economiá y Competividad; Netherlands Research School for Astronomy (NOVA); Netherlands Organisation for Scientific Research (NWO); Science and Technology Facilities Council; Swiss Space Office; Swedish National Space Board; UK Space Agency.

   We take this opportunity to thank the ESA JWST Project team and the NASA Goddard ISIM team for their capable technical support in the development of MIRI, its delivery and successful integration.

\end{acknowledgements}

\bibliography{aanda} 
\bibliographystyle{aa} 

\end{document}